\documentclass[letterpaper,english,aps,prb,floatfix,showpacs,amsfonts,amssymb,superscriptaddress]{revtex4}
\usepackage[T1]{fontenc}
\usepackage[latin1]{inputenc}
\usepackage{graphics}
\usepackage{amssymb}
\usepackage{amsmath}
\usepackage{overpic}

\usepackage{graphics}

\newcommand{\be}{\begin{equation}}
\newcommand{\ee}{\end{equation}}
\newcommand{\bea}{\begin{eqnarray}}
\newcommand{\eea}{\end{eqnarray}}

\def\a{\alpha}

\def\e{\varepsilon}
\def\d{\delta}

\def\m{\mu}
\def\l{\lambda}
\def\th{\theta}

\def\o{\omega}

\def\s{\sigma}

\def\D{\Delta}

\def\L{\Lambda}

\def\ra{\rightarrow}

\def\pll{\parallel}

\def\Ra{\Rightarrow}
\def\pd{\partial}
\def\nb{\nabla}
\def\bk{{\bf k}}
\def\bl{{\bf l}}
\def\br{{\bf r}}
\def\bj{{\bf j}}
\def\bq{{\bf q}}

\def\bA{{\bf A}}
\def\bB{{\bf B}}
\def\bH{{\bf H}}

\def\bJ{{\bf J}}
\def\bM{{\bf M}}
\def\bS{{\bf S}}

\def\DD{{\cal{D}}}
\def\GG{{\cal{G}}}

\def\nn{\nonumber}
\def\lb{\label}
\def\pref#1{(\ref{#1})}

\newcount\bozza \bozza=0
\ifnum\bozza=1
\newdimen\shift \shift=-2truecm
\def\lb#1{%
{\label{#1}\rlap{\kern\shift{$\scriptstyle#1$}}}}
\else\def\lb#1{\label{#1}} \fi

\begin{document}

\title{Beresinskii-Kosterlitz-Thouless transition
  within the sine-Gordon approach: the role of the 
vortex-core  energy}

\author{L.~Benfatto}
\affiliation{Institute for Complex Systems (ISC), CNR, U.O.S. Sapienza,\\ P.le A. Moro 2,
00185 Rome, Italy}
\affiliation{Department of Physics, Sapienza University of Rome, \\P.le A. Moro 2,
00185 Rome, Italy}

\author{C.~Castellani}
\affiliation{Department of Physics, Sapienza University of Rome, \\P.le A. Moro 2,
00185 Rome, Italy}
\affiliation{Institute for Complex Systems (ISC), CNR, U.O.S. Sapienza,\\ P.le A. Moro 2,
00185 Rome, Italy}

\author{T.~Giamarchi}
\affiliation{DMCP-MaNEP Universit\'e de Gen\`eve, Quai Ernest-Ansermet CH-1211
Gen\`eve 4, Switzerland}

\date{\today}

\begin{abstract}
  One of the most relevant manifestations of the
  Beresinskii-Kosterlitz-Thouless transition occurs in
  quasi-two-dimensional superconducting systems. The experimental
  advances made in the last decade in the investigation of
  superconducting phenomena in low-dimensional correlated electronic
  systems raised new questions on the nature of the BKT transitions in
  real materials. A general issue concerns the possible limitations of
  theoretical predictions based on the $XY$ model, that was studied as
  a paradigmatic example in the original formulation. Here we review
  the work we have done in revisiting the nature of
  the BKT transition within the general framework provided by the
  mapping into the sine-Gordon model. While this mapping was already
  known since long, we recently emphasized the advantages on such an
  approach to account for new variables in the BKT physics. One such variable
  is the energy needed to create the core of the vortex, that is
  fixed within the $XY$ model, while it attains substantially
  different values in real materials. This has interesting observable
  consequences, especially in the case when additional relevant
  perturbations are present, as a coupling between stacked
  two-dimensional superconducting layers or a finite magnetic
  field.
\end{abstract}


\maketitle

\section{Introduction}

Almost 40 years after the pioneering work by Beresinskii
\cite{berezinsky} and Kosterlitz and Thouless
\cite{kosterlitz_thouless,kosterlitz_renormalization_xy} the
Beresinskii-Kosterlitz-Thouless (BKT) transition is still the subject
of an intense experimental and theoretical research. On very general
grounds, what makes it so fascinating is the possibility of a phase
transition that is not driven by the explicit breaking of a given
symmetry, but is based on the emergence of a finite (and measurable) rigidity
of the system. The BKT transition was originally formulated within the context of the 
two-dimensional (2D) $XY$-model, which describes the exchange
interaction between classical two-component spins with fixed length
$S=1$: 
\be
\lb{xy}
H_{XY}=-J\sum_{\langle ij \rangle}\cos (\th_i-\th_j),
\ee
where $J$ is the spin-spin coupling constant and $\theta_i$ is the
angle that the $i$-th spin form with a given direction, and $i$ are
the sites of a square lattice. This model admits a continuous $U(1)$
symmetry (encoded in the transformation $\theta_i \ra \theta_i
+\chi$), that cannot be broken at finite temperature by an average
magnetization $\langle \bS \rangle=\langle e^{i\theta}\rangle$
different from zero because of the Mermin-Wagner theorem.
Nonetheless, the system can become ``stiff'' at low temperature with
respect to fluctuations of the $\theta$ variable, leading to a
power-law decay of the spin correlation functions, i.e. $\langle
e^{i(\theta(i)-\theta(0))} \rangle \simeq r^{-T/2\pi J}$, in contrast
to the exponential one expected in the truly disordered state. Such a
change of behavior cannot be ``smooth'', i.e. a phase transition
occurs in between, which appears to be controlled by the emergence of
vortex-like excitations. The original argument used if Ref.\
\onlinecite{kosterlitz_thouless} to capture the temperature scale of such
a transition is rather intuitive: in two dimensions both the energy
$E$ and the entropy $S$ of a single vortex excitation depend
logarithmically on the size $L$ of the system, so that the free energy
reads:
\be
\lb{free}
F=E-TS=(\pi J-2T)\ln \frac{L}{a}.
\ee
As a consequence, at temperatures larger than 
\be
\lb{tktapprox}
T_{BKT}\simeq \frac{\pi J}{2},
\ee
free vortices start to proliferate and destroy the quasi-long-ranged
order of the correlation functions.

As it was observed already in the original
papers\cite{kosterlitz_thouless,kosterlitz_renormalization_xy},
several physical phenomena are expected to belong to the same
universality class than the $XY$ model \pref{xy}, as for example the
superfluid transition in two dimensions. Later on it was realized that
the same idea can be applied also to superconducting (SC) thin
films\cite{doniach_films,halperin_ktfilms}, even though in the charged
superluid the logarithmic interaction between vortices is screened by
the supercurrents at a finite distance $\Lambda=\lambda^2/d$, where
$\lambda$ is the penetration depth of the magnetic field and $d$ is
the film thickness.\cite{pearl} In practice, for sufficiently thin films with
large disorder (so that $\lambda$ is also large) the electromagnetic screening effects
are weak enough to expect the occurrence of the BKT transition.

As a matter of fact, the case of SC thin films represented
one of the most studied applications of the BKT physics. In principle,
in this case one has also several possibilities to access
experimentally the specific signatures of the BKT physics.  For
example, by approaching the transition from below, the superfluid
density $n_s$ is expected to go to zero discontinuously at the BKT
temperature $T_{BKT}$, with an ``universal'' relation between
$n_s(T_{BKT})$ and $T_{BKT}$
itself\cite{nelson_univ_jump_prl77,review_minnaghen,giamarchi_book_1d},
that is the equivalent of the relation \pref{tktapprox}, since
$J\propto n_s$ (see Eq.\ \pref{defj} below).  Approaching
instead the transition from above one has in principle the possibility
to identify the BKT transition from the temperature dependence of the
SC fluctuations.  Indeed, in 2D the temperature
dependence of both the paraconductivity $\Delta \sigma\equiv
\sigma-\sigma_n$ and the diamagnetism $\chi_d$ is encoded in the
SC correlation length $\xi(T)$,\cite{halperin_ktfilms} which increases
approaching the transitions due to the increase of SC
fluctuations:
\be
\lb{scf} 
\Delta \sigma\propto \xi^2(T), \quad \chi_d\propto -\xi^2(T).
\ee
Within BKT theory $\xi(T)$ diverges exponentially at $T_{BKT}$ as
$\xi_{BKT}(T)\simeq a
e^{b/\sqrt{t}}$,\cite{kosterlitz_renormalization_xy,review_minnaghen}
where $t=T/T_{BKT}-1$, in contrast to the power-law $\xi_{GL}(T)\simeq
1/(T-T_c)$ expected within Ginzburg-Landau (GL)
theory\cite{varlamov_book}. As a consequence, by direct inspection of
the paraconductivity or diamagnetism near the transition one could
identify the occurrence of vortex fluctuations, which lead to an
exponential temperature dependence of the SC correlation length.

Quite interestingly, very direct measurements of the BKT universal
jump in the superfluid density of SC films became available only
recently,\cite{lemberger_InO,martinoli_films,lemberger_bilayer,pratap_apl10,pratap_bkt_prl11,lemberger_Bifilms_cm11}
due to the improvement of the experimental
techniques, triggered mostly by the investigation of high-temperature
superconductors in the late nineties. In particular, the use of the two-coil
mutual inductance technique\cite{lemberger_twocoils} turned out to be crucial to obtain the
absolute value of the superfluid density at zero temperature, which is
needed to compare the experimental data with the BKT
predictions. 
Recently a great deal of information has come also from Tera-hertz
spectroscopy\cite{armitage_films_prb07,armitage_InO_prb11,armitage_natphys11},
which probes the finite-frequency analogous of the superfluid-density
jump. At
the same time, in the last decade new 2D or quasi-2D SC
systems emerged where the BKT transition is expected to occur. To this
category belong for example the nanometer-thick layers of
SC electron systems formed at the interface between
artificial heterostructures made of insulating oxides as
LaAlO$_3$/SrTiO$_3$\cite{triscone_science07,triscone_nature08} or
LaTiO$_3$/SrTiO$_3$\cite{biscaras_natcomm10}, or at the liquid/solid
interface of field-effect transistors made with organic
electrolytes\cite{iwasa_natmat10}. A second remarkable example is
provided by layered 3D systems as cuprate high-temperature
superconductors, where the weak interlayer coupling makes it plausible
that at least in same regions of the phase diagram a BKT transition
could be at play\cite{emery,review_lee}. Even though the existence of
a BKT transition in bulk (3D) samples is still controversial, as we
shall discuss below, nonetheless in cuprates the proximity of the
SC phase to the Mott insulator leads to a large
penetration depth without need of introducing strong disorder, making
in principle thin-films of cuprate superconductors the best candidate
to study BKT
physics\cite{corson_nature99,martinoli_films,triscone_fieldeffect,lemberger_bilayer,lemberger_Bifilms_cm11,armitage_natphys11}. Recently,
much attention has been devoted also to artificial heterostructures
made of cuprates at different doping
level\cite{yuli_prl08,bozovic_science09}, where the observation of BKT
physics in the SC films is complicated even more by the proximity to
the non-SC correlated insulator.

A common characteristic of the cases mentioned above is that the BKT
transition is expected to occur in systems where electronic
correlations are not necessarily in the weak-coupling limit. This can
be due to the presence of strong disorder, as it is the case for thin
disordered films of conventional superconductors, to the artificial
spatial confinement, as in the SC interfaces, or to the
intrinsic nature of the system, as it occurs in cuprate
superconductors. As a consequence, several experimental results seem
to point towards a kind of ``unconventional'' BKT physics, which needs
to be addressed using a wider perspective than the one proposed in the
original formulation. A paradigmatic example of an apparent failure of
the standard BKT approach is posed by recent measurements of the
universal superfluid-density jump in
InO\cite{armitage_films_prb07,armitage_InO_prb11} and NbN\cite{pratap_apl10,pratap_bkt_prl11}
films. In a quasi-2D superconductor of thickness $d$ the energy scale corresponding to the
coupling $J$ of the $XY$ model \pref{xy} is the so-called superfluid
stiffness, which is connected to the (areal)
density of superfluid electrons $\rho_s^{2d}\equiv n_s d$, which in turn is measured via the
inverse penetration depth $\lambda$ of the magnetic field:
\be
\lb{defj}
J=\frac{\hbar^2\rho_s^{2d}}{4m}=\frac{\hbar^2 c^2 d}{16\pi e^2 \l^2 }.
\ee
This coupling has itself a bare temperature dependence $J(T)$ due to
the presence of quasiparticle excitations: however, one would expect
that at a temperature scale corresponding to the relation
\pref{tktapprox} free vortices start to proliferate, so that $n_s(T)$
jumps discontinuously to zero. In the experiments of Ref.\
\onlinecite{pratap_apl10,pratap_bkt_prl11} one can clearly see that as the film thickness
decreases $n_s(T)$ starts to deviate abruptly from its BCS
temperature dependence. However, such a deviation seems to occur at a
temperature {\em lower} than the one predicted by Eq.\ \pref{tktapprox}. The
same observation holds for finite-frequency measurements of $n_s(T)$
in InO films\cite{armitage_films_prb07,armitage_InO_prb11}, casting some
doubt on that ``universal'' relation between the superfluid density and
the critical temperature that is one of the hallmarks of the BKT
transition.

It is worth noting that while the measurement of the
superfluid-density behavior gives access to the most straightforward
manifestation of BKT physics, its
identification via SC fluctuations  is much more
subtle. Indeed, according to the general result \pref{scf}, one needs
in this case a controlled procedure to first extract the SC fluctuations
contribution, and then to fit it with the BKT expression for the
correlation length. Such a procedure is in general applied to the
paraconductivity, even though much care should be used to disentangle
GL from BKT fluctuations, as it has been discussed in a seminal paper
long ago by Halperin and Nelson\cite{halperin_ktfilms} (HN). Indeed,
 the BKT fit should be applied only in the region between  the BCS mean-field
temperature $T_c$ and the true $T_{BKT}$, that do not differ
considerably in thin films of conventional superconductors. In
contrast, recent applications to
paraconductivity measurements in thin films of cuprate
superconductors\cite{triscone_fieldeffect} or in SC
interfaces\cite{triscone_science07,schneider_finitesize09} seem to
suggest that in these systems the whole fluctuation regime above
$T_{BKT}$ is
dominated by BKT vortex fluctuations, and deviations only occur near
the transition because of finite-size effect. Also in this case, one
would like to distinguish unconventional effects due possibly to the
nature of the underlying system, from spurious results due to an 
incorrect application of BKT theory. As we discussed recently in Ref.\
[\onlinecite{benfatto_kt_interfaces}] a BKT fit of the paraconductivity must
be done taking into account from one side the existence of unavoidable 
constraints on the values of the fitting parameters, and from the other
side the existence of inhomogeneity on mesoscopic scales, that can
partly mask the occurrence of a sharp BKT transition. A very interesting
example of application of such a procedure has been recently provided by
NbN thin films\cite{pratap_bkt_prl11}, where the direct comparison between
superfluid-density data below $T_{BKT}$ and resistivity data above
$T_{BKT}$ provided a paradigmatic example of BKT transition in a real
system. 

The issue of the inhomogeneity emerges also within the context of
high-temperature cuprate superconductors. It is worth noting that in
the literature the discussion concerning the occurrence of BKT physics
in these layered anisotropic superconductors has been often associated
to a somehow related issue, i.e. the nature of the pseudogap state
above $T_c$. Indeed, despite the intense experimental and theoretical
research devoted to it, no consensus has been reached yet concerning
its origin, with two main lines of interpretation based either on a
preformed-pair scenario, or on the existence of a competing order,
associated to fluctuations in the particle-hole
channel\cite{review_lee}.  The preformed-pair scenario has in turn
triggered the attention on the role of SC phase fluctuations, which
are expected to be very soft in these materials having a small
superfluid density. Finally, the experimental observation of a large
Nerst effect\cite{ong_nerst_prb06,ong_natphys07,li_nerst_prb10} and
diamagnetism\cite{li_magn_epl05,li_nerst_prb10,rigamonti_prb10}well above $T_c$
has been used to support the notion that phase fluctuations have a
vortex-like character, as it is indeed the case within the BKT
picture. However, the overall interpretation of the experimental data
in terms of BKT physics is not so straightforward, despite several
theoretical attempts based both on the mapping into the Coulomb-gas
problem\cite{sondhi_kt,huse_prb08} or on numerical simulations for
the $XY$ model.\cite{podolsky_prl07} On the other hand, a
large Nerst effect arises also from the Fermi-surface reconstruction
associated to stripe order \cite{taillefer_nerst09}, or from ordinary
GL fluctuations,\cite{varlamov_prb11} as observed for example in thin films of conventional
superconductors\cite{lesueur_nerst}. Moreover, while the SC fluctuations
contribution to the diamagnetism seems to fit the BKT behavior of the
SC correlation length\cite{li_magn_epl05}, the paraconductivity shows
usually more direct evidence of GL
fluctuations\cite{leridon_prb07,rullier-albenque_prb11}.  Also the
issue of the superfluid-density jump is controversial: while it has
been clearly identified in very thin films\cite{lemberger_bilayer,armitage_natphys11,lemberger_Bifilms_cm11}, it
seems to be absent in bulk materials even for highly underdoped
samples\cite{broun_prl07}, where the low superfluid density and the
weak interplane coupling would make more plausible the presence of BKT
physics. The aim of this paper is not to give a detailed overview of
all the arguments in favor or against the occurrence of the BKT
transition in cuprates, but to focus on the correct identification of the
BKT signatures in a non-conventional quasi-2D superconductor. In general,
a layered system with very low in-plane superfluid density and very
weak interlayer coupling is one of the best candidates to observe those
signatures of BKT physics that we mentioned above, i.e. a rapid
downturn of the superfluid density coming from below $T_c$ and a
regime of vortex-like excitations above $T_c$.  However, since the
underlying superconductor is an unconventional one, the occurrence of
the BKT physics could be masked by other effects, making its
identification more subtle than in films of conventional
superconductors. We notice that addressing this issue does not solve
the more general problem of the nature of the pseudogap phase: for
example, to estimate the extension in temperature of the regime of BKT
fluctuations above $T_c$ one needs to consider physical ingredients
that are beyond the BKT problem addressed here. Nonetheless, a deeper
understanding of what could be the signatures of BKT physics can help
discriminating its occurrence or not in unconventional superconductors.

On the light of the above discussion, we will review in the present
manuscript the work we have done in the last years to investigate the
outcomes of the BKT transition in the presence of some additional ingredients
that influence its occurrence in real systems without invalidating the
basic physical picture behind it. The first ingredient that we will
consider is the role played by the vortex-core energy $\mu$, both in
quasi-2D systems and in layered ones. As we shall see, while in the
original formulation of BKT theory based on the $XY$ model \pref{xy}
$\mu$ is just a fixed constant times the coupling $J$, in real
materials $\mu/J$ can depend crucially on the microscopic nature of
the underlying system. Since it represents the energy scale needed to
create the core of the vortex, having ``cheap'' or ``expensive'' vortices can
influence in a non-trivial way the tendency of vortex formation below
and above the transition. Thus, while in 2D the critical behavior will
not change, in the layered 3D case the existence of expensive vortices
can move the vortex-unbinding transition away from the temperature
where it would occur in each (uncoupled) layer. 
The second aspect that we will discuss is the presence of
inhomogeneity on a mesoscopic scale. This issue is somehow related to
the effect of disorder on the BKT transition: however, instead of
considering a model of microscopic disorder, we will implement a
simpler approach where the spatial inhomogeneity of the superfluid
density can be mapped in a probability distribution of the possible
realizations of the superfluid-density values. This issue is in part
motivated by several 
experimental\cite{sacepe_stm_prl08,sacepe_stm_natcom10,sacepe_stm_natphys11,pratap_prl11,pratap_stm_cm11,yazdani_07}
and theoretical\cite{feigelman_review10,dubi_nature07,trivedi_cm10} suggestions that
inhomogeneity on a mesoscopic scale can occur both in
highly-disordered films of conventional
superconductors\cite{sacepe_stm_prl08,sacepe_stm_natcom10,sacepe_stm_natphys11,pratap_prl11,pratap_stm_cm11}
and in layered cuprate superconductors\cite{yazdani_07}, making then
timely to investigate its effect on the BKT transition.  Finally, we
will discuss the role of a finite external magnetic field, an issue
that is strictly related to the peculiarity of the BKT transition in a
charged superfluid. In this case the motivation comes in part from
recent experiments in cuprate superconductors, where anomalous
non-linear magnetization effects have been
reported\cite{li_magn_epl05,ong_natphys07,li_nerst_prb10} even for
those samples\cite{li_magn_epl05} where apparently clear signatures of
BKT physics appear. However, the main focus here is to establish a
clear theoretical framework to deal with this complicated problem,
based on the mapping into the sine-Gordon model. This last methodology
aspect will serve as a general guideline for the present review
paper. Indeed, we shall argue in the present manuscript that the 
mapping between the BKT transition in two
dimensions and the quantum phase transition within the sine-Gordon
model in one spatial dimension is the most powerful approach
to explore the outcomes of the BKT physics beyond the standard results
based on the original $XY$ model. In particular, such a mapping
provides us with a straightforward framework to explore the role of arbitrary values of
$\mu/J$, to include the effects of relevant perturbations (in the RG
sense) as the interlayer coupling or the magnetic field, and to
account at a basic level for the presence of inhomogeneities. 
Due to the several subtleties of such a mapping we shall 
first review the basic steps
of its derivation in Sec. \ref{mapping}, taking the point of view of the longitudinal {\em
  vs} transverse current decoupling in the $XY$ model. Once
established the general formalism and clarified the role of the
vortex-core energy we shall address in Sec. \ref{supdens}  the
consequences for the universal vs non-universal behavior of the superfluid density. 
In relation to the superfluid-density and paraconductivity behavior we 
shall give in Sec. \ref{inhomogeneity} a short account about the role of
inhomogeneity and its observation in recent experiments in several
systems. Finally, in Sec. \ref{magfield} we give a detailed derivation
of the sine-Gordon mapping in the presence of an external magnetic
field, that completes our overview on the approach to the BKT physics
in a real superconductor, and we discuss only briefly some physical
outcomes related to the previous Sections. The concluding remarks are
reported in Sec.\ \ref{conclusions}.

\section{Mapping on the sine-gordon model and the vortex-core energy}
\label{mapping}

As it is well known, the BKT transition occurs in three different
physical phenomena, that belong to the same universality class: the
vortex-unbinding transition within the $XY$ model \pref{xy}, the
charge-unbinding transition in the 2D Coulomb gas, and the quantum
metal-insulator transition in the 1D Luttinger liquid, as described by
the sine-Gordon model. In the first two cases we are dealing with a
{\em classical} model of point-like objects (vortices or charges)
interacting via a logarithmic potential, that becomes short-ranged
when the objects are free to move and to screen the interaction. In
the latter case we deal with a quantum 1D model, that becomes
effectively a 2D one at $T=0$ where dynamic degrees of freedom provide
the extra dimension. Even though all these analogies have been
reviewed several times in the literature (see for example
[\onlinecite{review_minnaghen,giamarchi_book_1d,nagaosa_book}] just to
mention a few references), nonetheless we will recall in this Section the
main steps of the mapping between the classical $XY$ model
and the quantum 1D sine-Gordon model. We shall take as a starting
point of view the separation between longitudinal and transverse
excitations of the phase, and we shall derive as an intermediate step
the mapping on the Coulomb-gas problem, to make the physical aspects
of the problem more evident. Instead in Sec. \ref{magfield}, where we
discuss the case of a finite magnetic field, we shall use a more
formal approach, that has however the great advantage to provide us
with an elegant and powerful formalism to discuss the case of a
superconductor embedded in an external electromagnetic potential.

As a starting point we shall consider a low-temperature limit for
the $XY$ model \pref{xy}, where one could expect
that the difference in angle between neighboring spins varies very slowly
on the scale $a$  of the lattice, so that one can approximate
$\theta_i-\theta_{i+\hat \delta}\approx a \pd \theta(\br) /\pd \hat\d$ where
$\theta(\br)$ is a smooth function and $\hat \delta=x,y$. Moreover, by
retaining the leading powers in the phase differences from the cosine in
Eq.\ \pref{xy}  we find that in the low-temperature phase the model reduces
to:
\be
\lb{h_j}
H_{XY}=\frac{J}{2}\int d\br (\nabla \theta)^2=\frac{J}{2}\int d\br \, \bj^2(\br),
\ee
where we introduced, in analogy with the case of the superfluid, a current
proportional to the phase gradient, $\bj= \nb \theta$. Because of the
smoothness assumption the approximation \pref{h_j} accounts only for
the longitudinal component $\bj_\pll$ of the current, while vortices,
i.e. singular configuration of the phase, 
can be associated to a transverse current component $\bj_\perp$. Indeed,
a vortical configuration for the phase $\theta_i$ of the $XY$ model
\pref{xy} corresponds to a 
non-vanishing circuitation of the phase
gradient along a closed line, that is nonzero only for a transverse
current $\bj_\perp$:
\be
\lb{vort}
\oint \bj \cdot d\ell=\int_S (\nb\times \bj)\cdot d{\bf s}=
\int_S (\nb\times \bj_\perp)\cdot d{\bf s}=2\pi \sum_i q_i,
\ee
where $q_i=\pm m$ is the vorticity of the $i$-th vortex, with $m$ integer. We can then
decompose in general the current of Eq.\ \pref{h_j} as
$\bj=\bj_\pll+\bj_\perp$, where $\nabla\times \bj_\pll=0$ and
$\nabla\cdot \bj_\perp=0$. One can easily see that the mixed terms
$\int d\br \bj_\pll\cdot \bj_\perp=0$ in Eq.\ \pref{h_j} vanish, so
that longitudinal and transverse degrees of freedom decouple
$H=H_\pll+H_\perp$, and we can focus on the term $H_\perp=\int d\br
\bj_\perp^2$ to describe the interaction between vortices.  By introducing a
scalar function $W$ the transverse current can be written as 
$\bj_\perp=\nb\times (\hat z W)=(\pd_y
W, -\pd_x W, 0)$, so that $\nb\times \bj_\perp=(0,0,-\nb^2
W)$ and inserting it into Eq.\ \pref{vort} we conclude that
$W$ must satisfy the equation:
\bea
\lb{eqphi}
\nb^2 W(\br)&=&-2\pi \rho(\br), \quad 
\rho(\br)=\sum_i q_i \d(\br-\br_i).
\eea
Eq.\ \pref{eqphi} is exactly the Poisson equation in 2D for the potential
$W$ generated by a distribution of point-like charges $q_i$ at the
positions $\br_i$. Its solution is in general:
\be
\lb{solphi}
W(\br)=2\pi\int d\br' V(\br-\br')\rho(\br'),
\ee
where $V(\br)$ is the
solution of the homogeneous equation for the Coulomb
potential in 2D
\be
\lb{poisson}
\nb^2 V(\br)=-\d(\br) \quad \Ra 
V(\br)=\int \frac{d\bk}{(2\pi)^2} \frac{e^{i\bk\cdot \br}}{\bk^2},
\ee
so that $V(r)\simeq -\ln r $ at large distances. 
Thanks to the results \pref{eqphi}-\pref{solphi} $H_\perp$ can be written as:
\bea
H_\perp&=&\frac{J}{2}\int d\br \bj_\perp^2=\frac{J}{2}\int d \br
(\nb\times \hat z  W)^2=
\frac{J}{2}\int
d\br (\nb W)^2=-\frac{J}{2}\int d\br W \nb^2 W=\nn\\
&=&\pi J \int d\br W(\br)\rho(\br)=2\pi^2 J \int d\br d\br' \rho(\br)
V(\br-\br') \rho(\br')=\nn\\
\lb{coulomb}
&=&2\pi^2 J\sum_{ij} q_i q_j V(\br_i-\br_j).
\eea
Eq. \pref{coulomb} expresses the electrostatic energy for a Coulomb gas
with charge density $\rho(\br)$, completing thus the analogy between the
system of vortices and the system of charges. The 2D Coulomb potential 
\pref{poisson} shows the characteristic infrared divergence which
reflects on the divergence of $V(\br=0)$ in the thermodynamic limit,
leading to the neutrality constraint for the gas. Indeed, by close inspection of
Eq.\ \pref{poisson} one sees that $V(\br=0)=\int_{1/L} dk (1/k)\sim
\ln L\ra  \infty$ as $L\ra \infty$.  If we separate this divergent term by defining
\be
V(\br)=V(0)+G(\br),
\ee
where now $G(r=0)=0$, in Eq.\ \pref{coulomb} we obtain:
\be 
\lb{div} 
2\pi^2 J\sum_{ij} q_i q_j [V(0)+G(\br_i-\br_j)]= 2\pi^2 J
V(0) \left(\sum_i q_i\right)^2 +2\pi^2 J\sum_{ij} q_i q_j G(\br_i-\br_j).
\ee 
Since the Boltzmann weight of each configuration is $e^{-\beta
  H_\perp}$, the divergence of $V(0)$ in the thermodynamic limit leads
to a vanishing contribution to the partition function, unless
\be
\lb{neutral}
\sum_i q_i=0,
\ee
which means that only neutral configurations are allowed. A second
consequence of the above discussion is that one should include a
cut-off for the smallest possible distance between two
vortices. Starting from the lattice $XY$ model \pref{xy} a natural
cut-off is provided by the lattice spacing $a$ in the original model,
which translates in the correlation length $\xi_0$ when applied to 
SC systems. The exact form of the function $G(\br)$ at
short distances defines then the energetic cost of a vortex on the
smallest scale of the system, i.e. the so-called vortex-core
energy. By computing $G(\br)$ on the lattice (see also Eq.\
\pref{vlattice} below), one can see that 
that at distances $r\geq a$ it can be well approximated as
\be
\lb{gapp}
G(r)\simeq-\frac{1}{4}-\frac{1}{2\pi}\ln\left(\frac{r}{a}\right).
\ee
Using the neutrality condition \pref{neutral}, the fact that $G(0)=0$ (so
that in the last term of Eq.\ \pref{div} one can use $i\neq j$) and
the form \pref{gapp}, Eq.\ \pref{coulomb} can be written as:
\bea
H_\perp&=&2\pi^2 J\sum_{i\neq j} q_i q_j G(\br_i-\br_j)=
-2\pi^2J \sum_{i\neq j}\left[
\frac{1}{4}+\frac{1}{2\pi}\ln\left(\frac{r_{ij}}{a}\right) \right]q_i
q_j=\nn\\
&=&-\frac{\pi^2 J}{2} \sum_{i\neq j} q_i q_j -\pi J 
\sum_{i\neq j} \ln\left(\frac{r_{ij}}{a}\right) q_i q_j=
\mu  \sum_i q_i^2 -\pi J 
\sum_{i\neq j} \ln\left(\frac{r_{ij}}{a}\right) q_i q_j
\eea
where we used $\sum_{i\neq j} q_i q_j=-\sum_i q_i^2$ from Eq.\
\pref{neutral} and we identified the vortex-core energy $\mu$ with
\be
\lb{defmu}
\mu=\mu_{XY}\equiv\frac{\pi^2 J}{2}
\ee
Finally, we can use the neutrality condition \pref{neutral} by imposing
that there are $n$ pairs of vortices of opposite vorticity. Moreover, we
shall consider in what follows only vortices of the lower vorticity $q_i=\pm
1$, so that $H_\perp$ reads:
\bea
\lb{hperp}
H_\perp&=& 2n\mu -\pi J 
\sum_{i\neq j}^{2n} \ln\left(\frac{r_{ij}}{a}\right) \e_i \e_j, \quad
\e_i=\pm 1.
\eea
Eq. \pref{hperp} describes the interaction between vortices in a given
configuration with $n$ vortex pairs. In the partition function of the
system we must consider all the possible values of $n$, taking into account
that interchanging the $n$ vortices with same vorticity gives the same
configuration (so one should divide by a factor $1/(n!)^2$). In conclusion
$Z$ reads:
\bea
Z&=&\sum_{n=1}^\infty \frac{1}{(n!)^2}\int d\br_1 \dots d\br_{2n}
e^{-\beta 2n\mu}e^{\pi \beta J 
\sum_{i\neq j}^{2n} \ln\left(\frac{r_{ij}}{a}\right) \e_i \e_j}\nn\\
\lb{zcg}
&=&\sum_{n=1}^\infty \frac{1}{(n!)^2} y^{2n}\int
d\br_1 \dots d\br_{2n}e^{ 
\sum_{i< j}^{2n} 2\pi \beta J\ln\left(\frac{r_{ij}}{a}\right) \e_i \e_j},
\eea
where we introduced the vortex fugacity
\be
y=e^{-\beta \mu}.
\ee

The explicit derivation of the partition function $Z$ has the great
advantage to allow us to recognize
immediately the analogy with the quantum sine-Gordon
model, defined by the Hamiltonian:
\be
\lb{sg2}
H_{sg}=\frac{v_s}{2\pi}\int_0^L dx \left[K(\pd_x\theta)^2+\frac{1}{K}
(\pd_x\phi)^2 - \frac{g_u}{a^2}\cos(2\phi)\right],
\ee
where\cite{giamarchi_book_1d} $\theta$ and $\pd_x\phi$
represent two canonically conjugated variables for a 1D chain of length
$L$, with $[\theta(x'),\pd_x \phi(x)]=i\pi\d(x'-x)$, $K$ is the
Luttinger-liquid (LL) parameter, $v_s$ the velocity of 1D fermions, and
$g_u$ is the strength of the sine-Gordon potential. In this
formulation, the role of the SC phase is played by the field
$\theta$. Indeed, when the coupling $g_u=0$ one can
integrate out the dual field $\phi$ to get the action
\be
\lb{s0}
S_0=\frac{K}{2\pi}\int dx d\tau \left[(\pd_x\theta)^2+(\pd_\tau\theta)^2
\right],
\ee
equivalent to the gradient expansion \pref{h_j} of the model
\pref{xy}, once considered that the
rescaled time  $\tau\rightarrow v_s\tau$ plays the role of the second
(classical) dimension.  The dual field $\phi$ describes instead the
transverse vortex-like excitations. This can be easily understood by
considering the quantum nature of the operators within the usual
language of the sine-Gordon model. Indeed, a vortex configuration requires that
$\oint \nabla \theta =\pm 2\pi$ over a closed loop, see Eq.\
\pref{vort} above.  Since $\phi$ is the
dual field of the phase $\theta$, a $2\pi$ kink in the field $\theta$ is
generated by the operator $e^{i2\phi}$,\cite{giamarchi_book_1d} i.e. by the sine-Gordon
potential in the Hamiltonian \pref{sg2}. More formally, one can show
that the partition function of the $\phi$ field in the sine-Gordon model 
corresponds to the \pref{zcg} derived above. To see this, let
us first of all integrate out the $\theta$ field in Eq.\ \pref{sg2},
to obtain 
\be
\lb{sg}
S_{SG}=\frac{1}{2\pi K}\int d\br (\nb \phi)^2-\frac{g}{\pi} \int d\br
\cos(2\phi).
\ee
The overall factor $Z_\pll=\Pi_{q>0}(1/\beta J\bq^2)$ due to the
integration of the $\theta$ field (corresponding to the longitudinal
excitations $Z_\pll=\int \DD \theta_\pll e^{-\beta H_\pll}$ in Eq.\ \pref{h_j} above) will be
omitted in what follows. 
We can treat the first term of the above action as the free
part $S_0$, and we can expand the exponential of the interacting part
in series of powers, so that
\be
\lb{zsg}
Z=\int \DD\phi e^{-S_0} \sum_{p=0}^\infty \frac{1}{p!}
d\br_1 \dots d\br_{p} \left(\frac{g}{\pi}\right)^p 
\cos(2\phi(\br_1)) \dots \cos(2\phi(\br_p)).
\ee
Here $\int \DD \phi$ is the functional integral over the $\phi$
 field. When we decompose each cosine term as
\be
\cos(2\phi(\br_i))=\frac{e^{i\phi(\br_i)}+e^{-i\phi(\br_i)}}{2}=
\sum_{\epsilon=\pm 1}\frac{e^{i\epsilon\phi(\br_i)}}{2},
\ee
we recognize that in Eq.\ \pref{zsg} we are left with the calculation of
average value of exponential functions over the Gaussian weight $S_0$,
i.e of factors
\be
\langle {e^{2i\sum_i\epsilon_i\phi(\br_i)}}\rangle=e^{2K 
\sum_{i< j} \ln\left(\frac{r_{ij}}{a}\right) \e_i \e_j}.
\ee
Here we used the well-known properties of Gaussian
integrals\cite{giamarchi_book_1d,nagaosa_book} that impose that the
above expectation value is non zero only for neutral configurations
$\sum_{i=1}^p\e_i=0$, in full analogy with the result found above for
the vortices. We then put again $p=2n$. Taking for instance $\e_1,
\dots \e_n=+1$ while $\e_{n+1}, \dots \e_{2n}=-1$ the combinatorial
prefactor $1=p!\equiv 1/(2n)!$ in Eq.\ \pref{zsg} should be multiplied
times the number $\left(2n \atop n\right)=(2n)!/(n!)^2$ of
possibilities to choose the $n$ positive $\e_i$ values over the $2n$
ones. Thus, Eq.\ \pref{zsg} reduces to:
\bea
\lb{zsgfin}
Z&=&\sum_{n=1}^\infty \frac{1}{(n!)^2} 
\left(\frac{g}{2\pi}\right)^{2n}
\int d\br_1 \dots d\br_{2n}
e^{ 2K\sum_{i< j}^{2n} \ln\left(\frac{r_{ij}}{a}\right) \e_i \e_j}.
\eea
By comparing Eq.\ \pref{zcg} and Eq.\ \pref{zsgfin} we see that the vortex
problem (as well as the Coulomb-gas problem) is fully mapped into the
sine-Gordon model, provided that we identify:
\bea
\lb{defk}
K&=&\frac{\pi J}{T},\\
\lb{defg}
g&=&2\pi e^{-\beta\mu}.
\eea

As it is clear from the above derivation, within the $XY$ model there exists
a precise relation \pref{defmu} between the value of the vortex-core energy $\mu$ and
the value of the superfluid coupling $J$. This is somehow a natural
consequence of the fact that the $XY$ model \pref{xy} has only {\em one}
coupling constant, $J$. Thus, when deriving the mapping on the continuum
Coulomb-gas problem \pref{hperp} $\mu$ is fixed by the short length-scale
interaction, that fixes the behavior of $G(r)$ in Eq.\ \pref{gapp} and
consequently the vortex-core energy value \pref{defmu}. In contrast, within
 the sine-Gordon language $\mu$ is determined by the value of the
 interaction $g$ for the model \pref{sg}, which can attain in principle
 arbitrary values. Thus, such a mapping is the more suitable one to investigate
 situations where $\mu$ actually deviates from the $XY$-model value, as it
 is suggested by the physics of various systems.

A typical example is provided by the case of ordinary films of
superconductors, where usually the BCS approximation -and its dirty-limit
version- reproduces quite well the values of the SC quantities, like the gap
and the transition temperature.\cite{pratap_apl10,pratap_bkt_prl11} 
In this case, one would also expect that
$\mu$ has a precise physical analogous with the loss in condensation energy
within a vortex core of size of the order of the coherence length
$\xi_0$, 
\be
\lb{econd}
\mu=\pi \xi_0^2 \e_{cond}
\ee
where $\e_{cond}$ is the condensation-energy density.  In the clean case Eq.\ \pref{econd}
can be expressed in terms of $J_s$ by means of the BCS relations for 
$\e_{cond}$ and $\xi_0$. Indeed, since
$\e_{cond}=dN(0)\Delta^2/2$, where $N(0)$ is the density of states at the
Fermi level and $\Delta$ is the BCS gap, and 
$\xi_0=\xi_{BCS}=\hbar v_F/\pi \Delta$, where $v_F$ is the Fermi velocity, and  assuming
that $n_s=n$ at $T=0$, where $n=2N(0)v_F^2 m/3$, one has
\be
\lb{mu}
\mu_{BCS}= \frac{\pi \hbar^2 n_s d}{4 m} \frac{3}{\pi^2}=\pi J_s \frac{3}{\pi^2}
\simeq 0.95 J_s,
\ee
so that it is quite smaller than in the $XY$-model case
\pref{defmu}. This can have profound physical consequences on the
manifestation of the BKT signatures in real materials, as we shall discuss
in more details in the next Section.

\section{The universal jump of the superfluid density}
\label{supdens}

The above derivation of the mapping between the $XY$ model and the
sine-Gordon model can be used directly to identify the two relevant
running couplings $K$ and $g$ that must be considered under
renormalization group. The coupling $K$ \pref{defk} is connected to
the superfluid behavior of the system, thanks to the identification
\pref{defj} of $J$ with the superfluid stiffness of the system. Such
an identification relies on the analogy between Eq.\ \pref{h_j} and
the kinetic energy of a superfluid. As a consequence, when also vortex
excitations are present the physical $J_s$ must account also for
vortex-antivortex pairs at short
distances.\cite{nelson_univ_jump_prl77} This physical picture has a
precise correspondence on the values of the coupling constants under RG
flow, whose well-known equations
are\cite{kosterlitz_renormalization_xy,review_minnaghen,giamarchi_book_1d}:
\bea
\lb{eqk} 
\frac{dK}{d\ell}&=&-K^2g^2,\\ 
\lb{eqg} 
\frac{dg}{d\ell}&=&(2-K)g.
\eea
where $\ell=\ln a/a_0$ is the rescaled length scale. 
The superfluid stiffness is then identified by the limiting value of $K$
as one goes to large distances, i.e.
\be
\lb{jlim}
J_s\equiv \frac{T K(\ell\ra \infty)}{\pi}.
\ee
Even though the behavior of the RG equations \pref{eqk}-\pref{eqg} has
been described at length in several papers, we want to recall here the
basic ingredients needed to describe the BKT transition.  There are
two main regimes: for $K\gtrsim 2$ the r.h.s. of Eq.\ \pref{eqg} is
negative, so that $g\ra 0$ and $K$ tends to a finite value $\ra K^*$
that determines the physical stiffness $J_s$, according to Eq.\
\pref{jlim}. Instead for $K\lesssim 2$ the vortex fugacity grows under
RG flow, $K$ in Eq.\ \pref{eqk} scales to zero, and $J_s=0$. The BKT
transition temperature is defined as the highest value of $T$ such
that $K$ flows to a finite value. This occurs at the fixed point
$K=2,g=0$, so that at the transition one always have
\begin{equation}
\lb{jump}
K(\ell\ra\infty,T_{BKT})=2, \Ra \frac{\pi J_s(T_{BKT})}{T_{BKT}} = 2.
\end{equation}
As soon as one goes to temperatures larger than $T_{BKT}$ $K\ra 0$, so
also $J_s\ra0$. As a result, one finds $J(T_{BKT}^+)=0$, i.e. the
superfluid density jumps discontinuously to zero right above the
transition.  The equation \pref{jump} describes the so-called
universal relation between the transition temperature $T_{BKT}$ and
the value of the superfluid stiffness $J_s$ at the transition, and
represents a more refined version of the relation \pref{tktapprox}
based on the balance between the the energy and the entropy of a
single-vortex configuration.

\begin{figure}
\begin{center}
\parbox{2.1in}{\includegraphics[width=6cm]{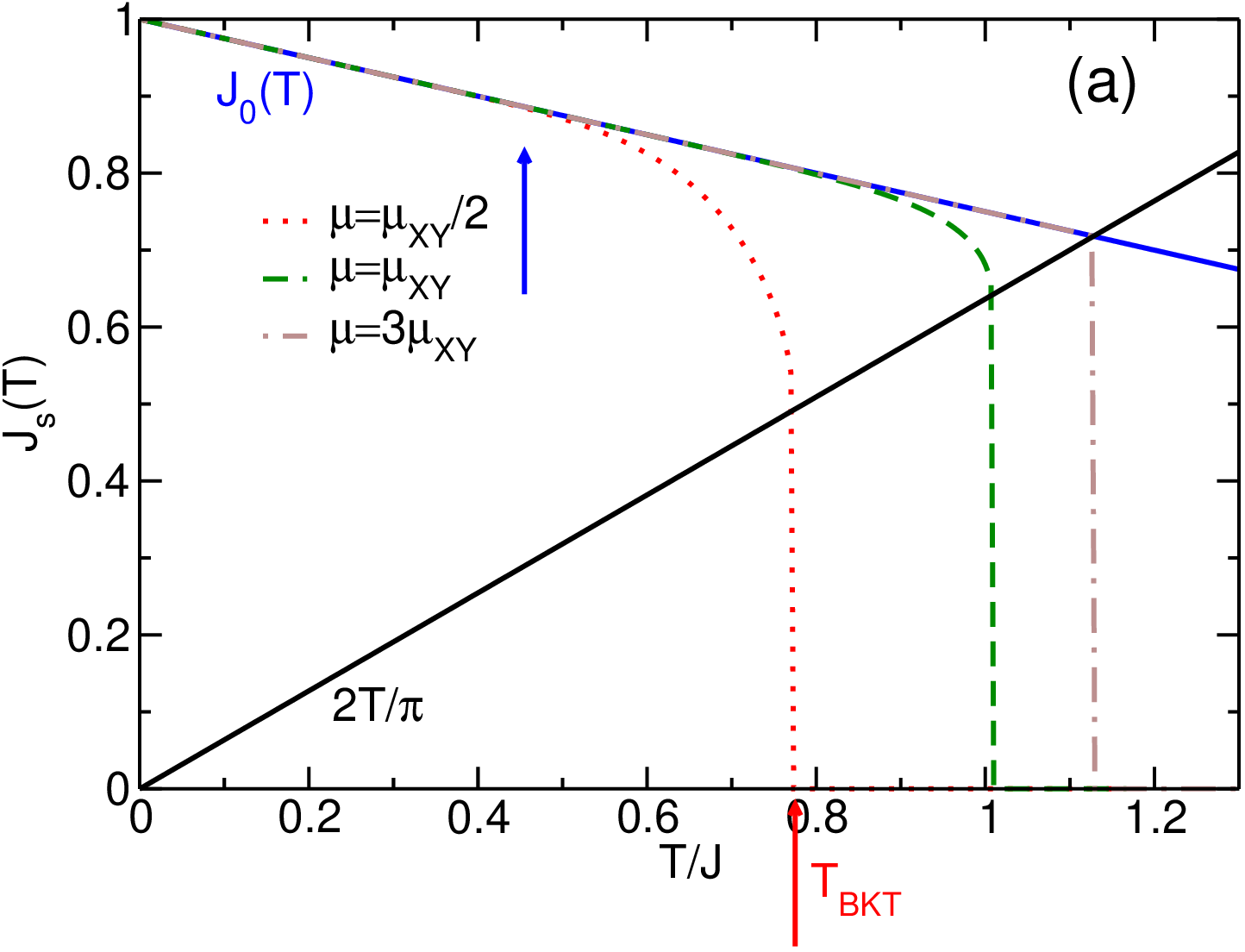}}
\hspace*{30pt}
\parbox{2.1in}{\includegraphics[width=6cm]{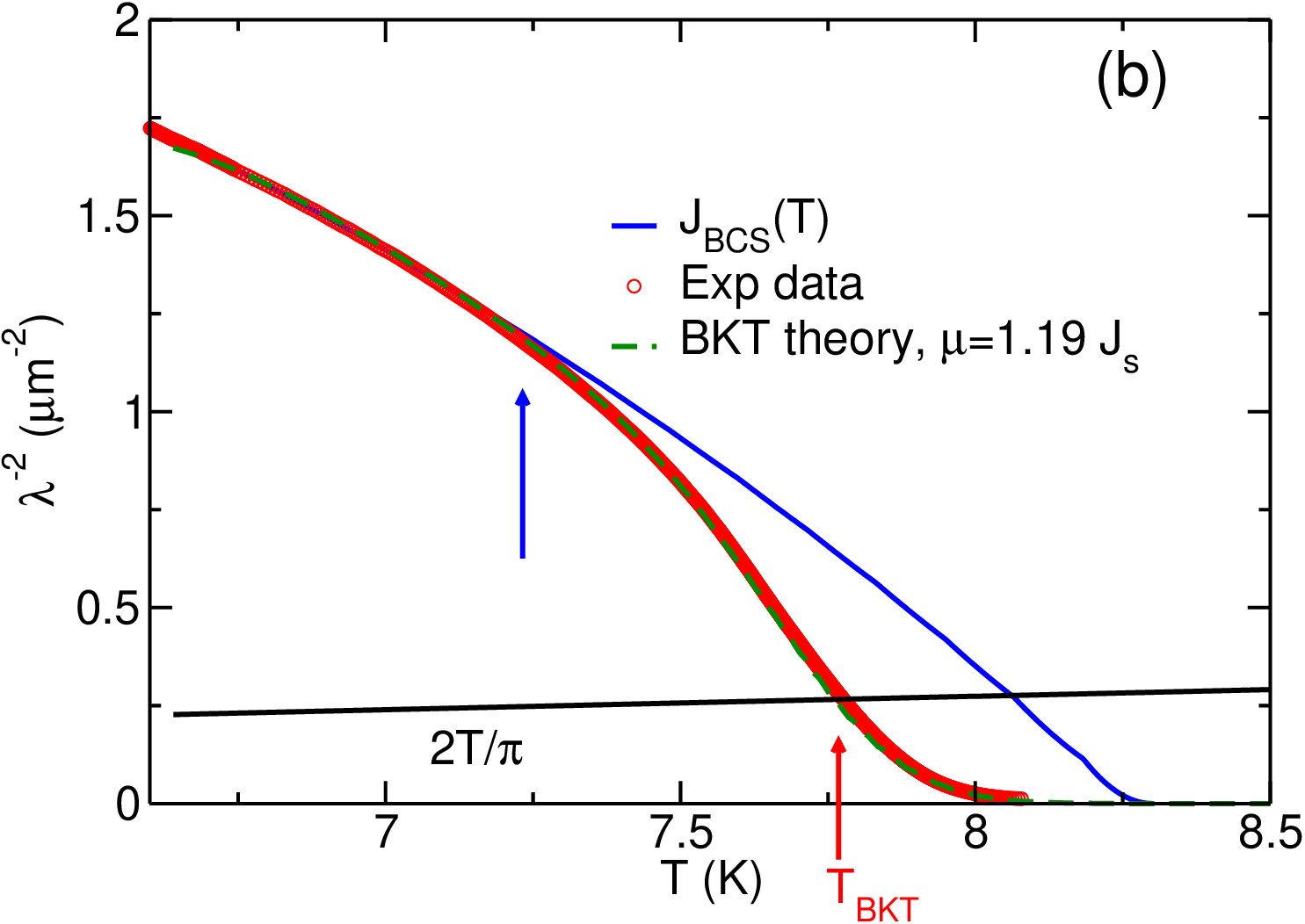}}
\caption{(a) Temperature dependence of the superfluid density for different
  values of the ratio $\mu/J$, measured in units of the value \pref{defmu}
  it has within the XY model. Notice that for small $\mu$ values the
  deviation from $J_0(T)$ starts much before (blue arrow) than the
  temperature where the universal jump
  (red arrow) occurs. This behavior is indeed observed in thin NbN
  films, panel (b). Here we report data from Ref.\
  [\onlinecite{pratap_bkt_prl11}] along with the theoretical BKT fit. Notice that
  the jump here is further smeared out by the inhomogeneity, see
  Sec.\ \ref{inhomogeneity} below.  }
\label{fig_mu}
\end{center}
\end{figure}

It should be noticed that the BKT RG equations account only for the effect
of vortex excitations, so that any other excitation that contributes to the
depletion of the superfluid stiffness must be introduced by hand in the
initial values of the running couplings. For example, in real
superconductors there are also quasiparticle excitations, while in the XY
model there are also longitudinal phase fluctuations, that give rise to a
linear depletion to the superfluid stiffness $J_0(T)=J(1-T/4J)$. As a
consequence the hand-waving argument usually adopted in the literature to
estimate $T_{BKT}$ in a system that is expected to have a BKT transition is
to look for the intersection between the universal line $2 T/\pi$ and the
$J(T)$ expected from the remaining excitations except than
vortices. However, such a procedure can only be approximate, since in the
relation \pref{jump} the temperature dependence of $J_s(T)$ is determined
also by the presence of bound vortex-antivortex pairs, which can renormalize
$J_s$ already {\em below} $T_{BKT}$. This effect is usually negligible when
$\mu$ is large, as it is the case for superfluid films\cite{bishop_prb80}
or within the standard $XY$ model. However, as $\mu$ decreases the
renormalization of $J_s$ due to bound vortex pairs increases, and
consequently $T_{BKT}$ is further reduced with respect to the
mean-field critical temperature $T_c$. As an
example we show in Fig.\ \ref{fig_mu} the behavior of $J_s(T)$ using a bare
temperature dependence as in the $XY$ model and switching the vortex-core
energy from the value \pref{defmu} to values smaller or larger. As one can
see, for decreasing $\mu$ the effect of bound vortex-antivortex pairs below
$T_{BKT}$ is significantly larger, moving back the transition temperature
to smaller values. The very same effect has been recently observed in thin
films of NbN\cite{pratap_apl10,pratap_bkt_prl11}, where it has been shown
experimentally that the deviation of $J_s(T)$ from the BCS curve starts
significantly below the transition temperature. Interestingly, the direct
comparison between the experimental $J_s(T)$ and the results of the RG
equations allowed the authors to show that the vortex-core energy in this
system attains indeed a value of the order of the BCS estimate \pref{mu}.

It must be emphasized that the case of thin NbN films must be seen as a
paradigmatic example of manifestation of the {\em universal} relation
\pref{jump}, despite the fact that $T_{BKT}$ is lower than expected
for a standard view based on the $XY$-model results. Indeed, in this system once
that the {\em renormalized} stiffness is of the order of $2T/\pi$ the
transition actually occurs. A different behavior is instead observed in the
case we discussed recently\cite{benfatto_kt_rhos} within the context of cuprate superconductors, that
can be well modeled as weakly coupled 2D layers. In this case, an
additional energy scale exists, i.e. the Josephson coupling $J_c$
between layers, which is also a relevant coupling under RG flow, so that 
a vortex-core energy different from the
$XY$-model value can lead to a qualitatively different behavior. We notice that the case
of weakly-coupled layered superconductors has been widely investigated in the past within
the framework of a layered version of the $XY$-model \pref{xy}. In this
case the presence of a finite interlayer coupling $J_c\ll J$ cuts off the
logarithmic divergence of the in-plane vortex potential at scales $\sim
a/\sqrt{\eta}$,\cite{cataudella} where $\eta=J_c/J$, so that the
superfluid phase persists above $T_{BKT}$, with $T_c$ at most few percent
larger than $T_{BKT}$.  \cite{chattopadhyay,friesen,pierson} As far as the
superfluid density is concerned, there is some theoretical\cite{chattopadhyay} and
numerical\cite{minnaghen_mc_xymodel} evidence that even for moderate anisotropy the
universal jump at $T_{BKT}$ is replaced by a rapid downturn of $\rho_s(T)$
at a temperature $T_d$ larger than the $T_{BKT}$ of each layer, but still very
near to it $T_d\simeq T_{KT}$. Once more this result must be
seen as an indication that the (only) scale $J$ dominates the problem: the
result can be different when $\mu$ is allowed to vary, making the
competition with the interlayer coupling more subtle. 

The analysis of the more general case has been done in a very convenient
way in Ref.\ \onlinecite{benfatto_kt_rhos}
within the framework of the sine-Gordon model \pref{sg2}, that must be suitable
extended to include the interlayer coupling. As we said, in the
sine-Gordon model \pref{sg2} the variable $\theta$ represents the SC
phase. Since the phases in neighboring layers are coupled via a
Josephson-coupling like interaction, the most natural assumption  is
an additional cosine term of strength $J_c$ for the interlayer phase difference, which
translates in an interchain
hopping term in the language of the 1D quantum model. A
similar model has been also derived recently in Ref.\
[\onlinecite{nandori_jpcm07}] by using as the starting point the Lawrence-Doniach model for the
layered superconductor. The full Hamiltonian that we consider is:\cite{benfatto_kt_rhos}
\be
\lb{ham}
H=\sum_m H_{sg}[\phi_m,\theta_m]
-\frac{v_sg_{J_c}}{2\pi a^2}\sum_{\langle m,m'\rangle}
\int_0^L dx \cos[\theta_m-\theta_{m'}],
\ee
where $m$ is the layer (chain) index and $g_{J_c}\equiv\pi J_c/{T}$. In
Ref.\ \onlinecite{benfatto_kt_rhos} we derived the perturbative RG equations
for the couplings of the model \pref{ham} by means of the operator product
expansion, in close analogy with the analysis of Ref.\
[\onlinecite{ho_deconfinement_short,cazalilla_deconfinement_long}] for
the multichain problem. Under RG
flow an additional coupling $g_\perp$ between the phase in neighboring
layers is generated:
\be
\lb{gf}
\frac{g_\perp}{2\pi}\sum_{\langle m,m'\rangle} \int dx \left[
-K(\pd_x\theta_m)(\pd_x\theta_{m'})+\frac{1}{K}
(\pd_x\phi_m)(\pd_x\phi_{m'})\right],
\ee
which contributes to the superfluid stiffness $K_s$, defined as usual as the
second-order derivative of the free energy with respect an infinitesimal
twist $\delta$ of the phase, $\pd_x \theta_m\rightarrow
\pd_x\theta_m-\delta$. Thus, one immediately sees that Eq.\ \pref{gf}
represents an interlayer current-current term, which contributes to the
in-plane stiffness $J_s$ as:
\be
\lb{defks}
K_s=K-nKg_\perp,  \quad J_s=\frac{K_s(\ell\ra\infty) T}{\pi},
\ee
where $n=2$ corresponds to the number of nearest-neighbors layers. The
full set of RG equations for the couplings $K,K_s,g_u,g_{J_c}$ reads:
\bea
\lb{eqk2}
\frac{dK}{d\ell}&=&2g_J^2-K^2g_u^2,\\
\lb{eqgu}
\frac{dg_u}{d\ell}&=&(2-K)g_u,\\
\lb{eqks}
\frac{dK_s}{d\ell}&=&-g_u^2K_s^2,\\
\lb{eqgj}
\frac{dg_{J_c}}{d\ell}&=&\left(2-\frac{1}{4K}-\frac{K_s}{4K^2}\right)g_{J_c}.
\eea
Observe that for $g_{J_c}=0$ the first two
equations reduce to the standard ones \pref{defk}-\pref{defg} of the BKT
transition, and $K_s$ coincides with $K$.
Instead, as an initial value $g_{J_c}\neq 0$ is considered, the interlayer 
coupling increases under RG,\cite{chattopadhyay,friesen,pierson}
contributing to stabilize the $K$ parameter in Eq.\ \pref{eqk2}, with
a consequent slowing down of the increase of the $g_u$ coupling in
Eq.\ \pref{eqgu}. As in the pure 2D case, in
the regime where $K$ goes to zero $g_u$ increases, see Eq.\ \pref{eqgu},
and vortices proliferate. However, in contrast to the single-layer case
where $g_u$ becomes always relevant near $K\simeq 2$, here thanks to the
$g_{J_c}$ term in Eq.\ \pref{eqk2}, $K$ can become smaller than 2 before than $g_u$
starts to increase. Since $K_s$ is controlled by the $g_u$
coupling alone, this means that the system remains superfluid in
a range of temperature above $T_{BKT}$ that depends on the competition
between vortices ($g_u$) and interlayer coupling ($g_{J_c}$). 

\begin{figure}
\begin{center}
\parbox{2.1in}{\includegraphics[height=5cm]{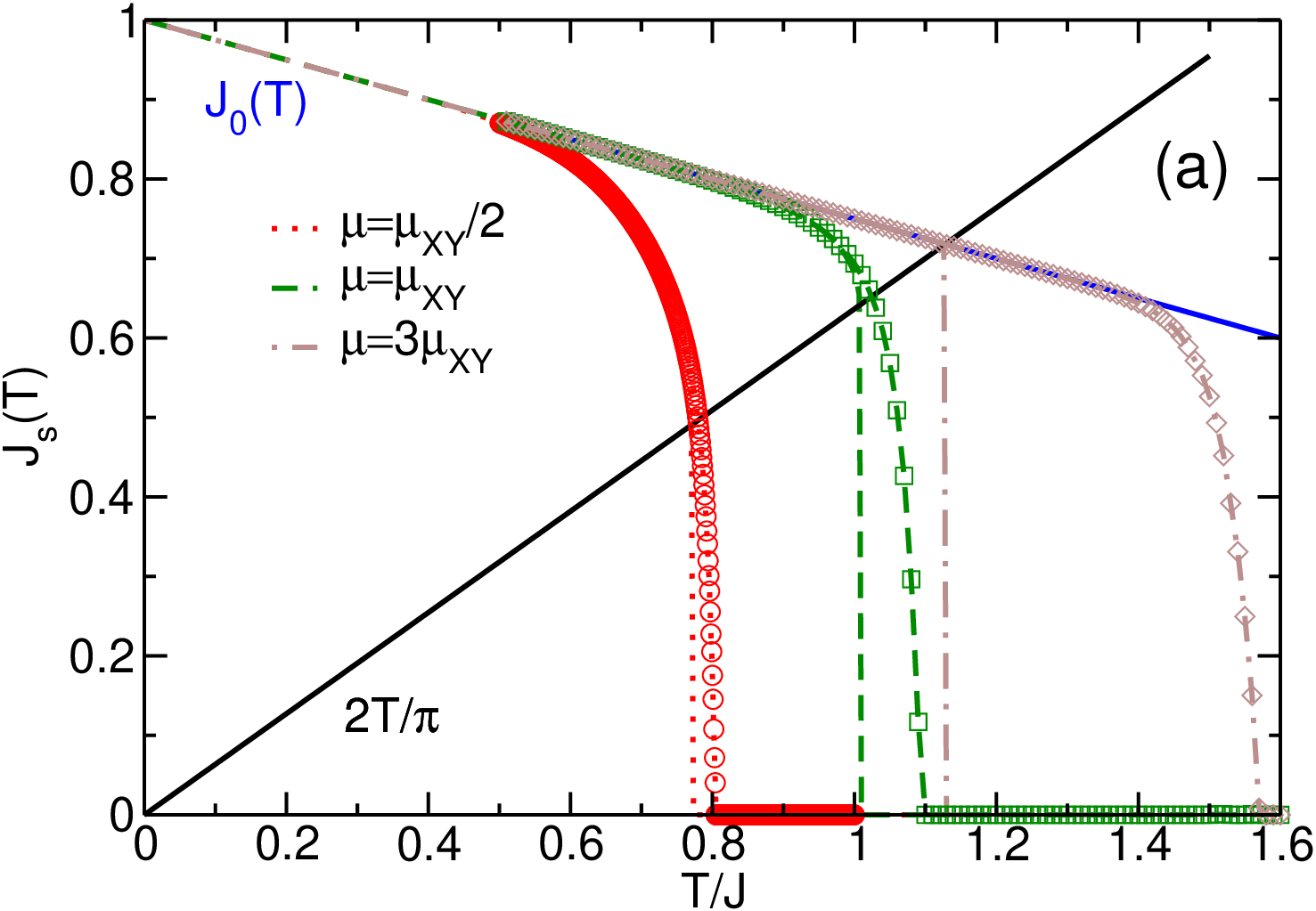}}
\hspace*{30pt}
\parbox{2.1in}{\includegraphics[height=5cm]{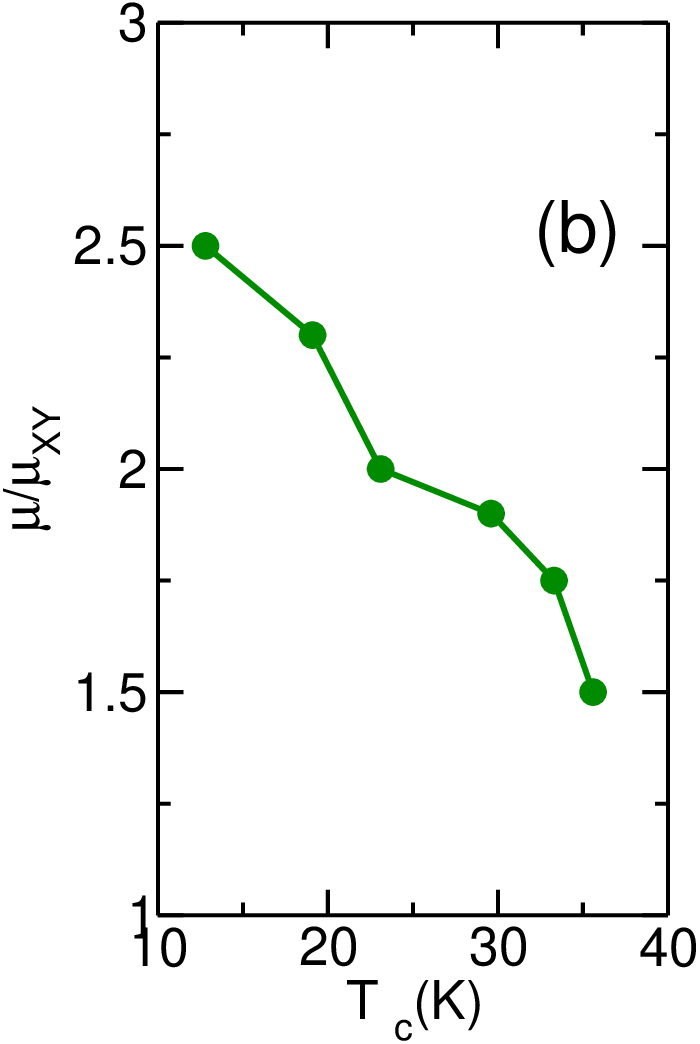}}
\end{center}
\caption{(a) Temperature dependence of the superfluid stiffness $J_s(T)$
  in the layered 3D case (symbols), taken from Ref.\
  [\onlinecite{benfatto_kt_rhos}]. Here $J_0(T)=J(1-T/4J)$
  (solid line) mimics the effect of longitudinal excitations within
  the $XY$ model.  It is also shown for comparison the 2D case (lines),
  see Fig.\ \ref{fig_mu}. For $\mu\leq\mu_{XY}$ $J_s(T)=0$ at
  $T_d\simeq T_{BKT}$. As $\mu$ increases $T_d$ increases as well, so
  that at $T_{BKT}$ no signature is observed in $J_s(T)$ of the jump
  present in the 2D case. (b) Vortex-core energy in units of
 $\mu_{XY}$ as a function of the critical temperature of several
 YBa$_{2}$Cu$_{3}$O$_{6+x}$ samples, as derived in Ref.\ [\onlinecite{benfatto_kt_bilayer}].}
\label{fig_js}
\end{figure}

The resulting
temperature dependence of the superfluid stiffness $J_s$ for different
values of the vortex-core energy is reported in
Fig.\ \ref{fig_js}a, where we added for the sake of completeness a temperature dependence of
the bare coupling $J_{0}(T)=J(1-T/4J)$, which mimics the effect
of long-wavelength phase fluctuations in the $XY$ model
\pref{xy}. Here the lines represent the pure 2D case (already shown in
Fig.\ \ref{fig_mu}a above), while the symbols are the result of the RG
equations \pref{eqk}-\pref{eqgj} for a fixed value $J_c/J=10^{-4}$ of the
interlayer coupling. As one can see, as soon as a finite interlayer coupling
is switched on, the jump of $J_s(T)$ at $T_{BKT}$ disappears and it is
replaced by a rapid bending of $J_s(T)$ at some temperature $T_d$. However,
while for $\mu\leq\mu_{XY}$ $T_d$ coincides essentially with $T_{KT}$, for
a larger vortex-core energy $T_d$ rapidly increases and deviates
significantly from the temperature scale where $J_s(T)$ intersects the
universal line $2T/\pi$. 

These results offer a possible interpretation of the measurements of
superfluid density reported in strongly-underdoped samples of 
YBa$_{2}$Cu$_{3}$O$_{6+x}$\cite{broun_prl07}. These systems are exactly the ones where a BKT
behavior could be expected: they have a very low in-plane superfluid
density and a large ($\eta\sim 10^{-4}$ ) anisotropy of the coupling
constants. However, the measured $J_s(T)$ goes
smoothly across the $T_{BKT}$ estimated from the 2D relation
\pref{jump}. In view of the above discussion, such a behavior does not
automatically rules out the possibility that any BKT physics is at play:
indeed, if the vortex-core energy is larger than in the $XY$ model the
transition can move away from the universal 2D case. Moreover, the effect
of inhomogeneity can further round off the downturn induced by vortex
proliferation (see next section), making it hardly
visible in the experiments. In analogy with the case of 
conventional superconductors discussed above, a hint on the realistic value of the vortex-core energy
in cuprate systems can be inferred by the direct comparison between
superfluid-density measurements and the theoretical prediction. We carried
out such an analysis in the case of bilayer films of underdoped
YBCO\cite{lemberger_bilayer} for several doping values, showing that $\mu$
attains always values larger than in the $XY$ model, with
$1.5<\mu/\mu_{XY}<3$, see Fig.\ \ref{fig_js}b. Moreover, both $\mu$ and the inhomogeneity 
increase as the system gets underdoped, supporting further the
interpretation that a similar effect can be at play also in the data of
Ref.\ [\onlinecite{broun_prl07}]. Interestingly, a very similar trend has been
reported recently in NbN systems as disorder increases\cite{pratap_bkt_prl11}, and it has been
interpreted by the authors as an effect of the increasing separation
between the energy scales associated to local pairing and superfluid
stiffness as disorder increases. These results suggest that also disorder
and inhomogeneity play a crucial role in the understanding of the BKT
transition, as we shall discuss in the next section.

\section{Inhomogeneity}
\label{inhomogeneity}

\begin{figure}[ht]
\begin{center}
\includegraphics[width=8cm]{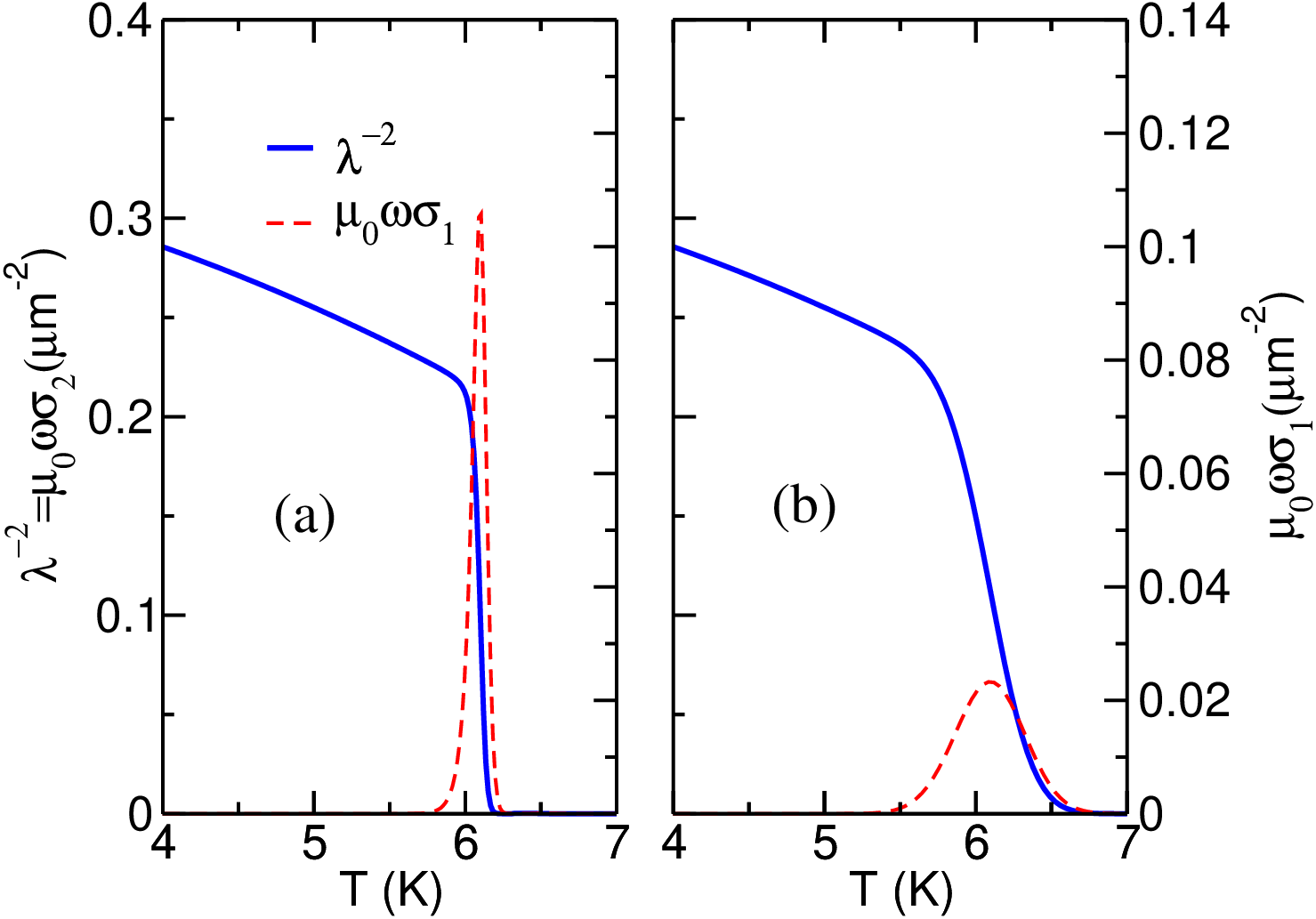}
\end{center}
\caption{Role of inhomogeneity, from Ref.\
  [\onlinecite{benfatto_kt_bilayer}]. (a)
 $1/\l^2$ and $\m_0\o\s_1$ evaluated at $\o=50$ KHz for a single 
 $\bar J(T)$ curve using parameter values appropriate for YBa$_{2}$Cu$_{3}$O$_{6+x}$
 films from Ref.\ [\onlinecite{lemberger_bilayer}]. Here $\mu=3\mu_{XY}$. The
 finite frequency leads to a sharp but continuous decrease of
 $1/\l^2$ across $T_{BKT}$, along with a peak in $\s_1$. (b) $1/\l^2$ and
 $\m_0\o\s_1$ evaluated at finite frequency taking into account the
 presence of inhomogeneities, as explained in the text.}
\label{fig-cond}
\end{figure}

As we mentioned in the introduction, detailed measurements of
superfluid density in thin films of superconductors became available only
recently thanks to the efficient implementation of the two-coils mutual
inductance technique, which gives access to the absolute value of the penetration depth of thin SC
materials. Quite interestingly, in all the cases reported so far in the
literature, concerning both conventional
superconductors\cite{pratap_apl10,pratap_bkt_prl11} and high-temperature
superconductors\cite{martinoli_films,lemberger_bilayer,bozovic_science09,armitage_natphys11,lemberger_Bifilms_cm11} the BKT
transition is never really sharp. At first sight one could wonder if any
finite-size effect is at play, as due to several factors: (i) the
existence of a finite screening length $\Lambda\simeq 2\lambda^2/d$ due to
the supercurrents in a charged superfluid\cite{pearl}; (ii) the finite
dimension of the system or (iii) the finite length $r_\omega$ intrinsically associated
to the probe, which uses an ac field at a typical frequency $\omega$ of the
order of the KHz. In all the above cases one should cut-off the RG
equations \pref{eqk}-\pref{eqg} at a finite scale $\ell_{max}=\ln
L_{max}/a_0$, leading to a rounding of the abrupt jump of the stiffness
at $T_{BKT}$. However, in practice such rounding effects are hardly
visible, since the decrease of $K(\ell)$ at $T>T_{BKT}$ is very fast,
leading to visible rounding effect only for very short cut-off length
scales of order of $\ell\simeq 2-3$. In real systems the cut-off length
scales are usually much larger: for example, both in the case of the NbN films of Ref.\
[\onlinecite{pratap_apl10}] and in the case of thin YBCO films from Ref.\ [\onlinecite{lemberger_bilayer}]  
$1/\lambda^2$ is or order of about 1 $\m$m$^{-2}$ near the transition and
$d$ is of the order of 1 nm, so that 
the Pearl length $\Lambda$ is of the order of 1mm, i.e. comparable to the
system size, leading to $\ell_{max}$ around $10$ ($a_o$ is of the order of
the coherence length $\xi_0\sim 1$ nm), which is practically an
infinite cut-off for the RG. At the same time $r_\omega=\sqrt{14 D/\o}$
is the maximum length probed by the oscillating field, where $D\sim
\hbar/m=10^{16}$ \AA$^2$/s is the diffusion constant of vortices and
$\omega\simeq 50$ KHz is the frequency of the measurements\cite{lemberger_bilayer,pratap_apl10},  giving again
a large cut-off scale $r_\omega\sim 0.1$ mm and no visible rounding effect,
see Fig.\ \ref{fig-cond}a. It is worth noting that in the
case of the experiments of Ref.\ \onlinecite{lemberger_bilayer} a second
indication of the existence of pronounced rounding effects come from the
experimental observation of a wide peak in the imaginary part of the
conductivity around $T_{BKT}$. Following the dynamical BKT theory of Ambegaokar et
al.\cite{ambegaokar_fin_freq} and Halperin and
Nelson\cite{halperin_ktfilms} such a peak is due to the bound- and
free-vortex excitations contribution to the complex
dielectric constant $\e(\o)=\e^1+i\e^2$ which appears in the 
complex conductivity $\s=\s_1+i\s_2$ at a finite
frequency $\o$:
\be
\lb{defs}
\s(\o)=-\frac{1}{\l^2e^2\m_0}\frac{1}{i\o \e(\o)}.
\ee
Here $\mu_0$ is the vacuum permettivity and we used MKS units as in
Ref.s\ [\onlinecite{benfatto_kt_bilayer,lemberger_bilayer}].
As in the case of the superfluid-density jump, we estimated\cite{benfatto_kt_bilayer} the width
$\Delta T_\omega$ in $\sigma_1(\omega)$ due to the finite frequency
$\omega$ of the measurements and we showed that it is expected to be much
smaller than the experimental observations in Ref.\
\onlinecite{lemberger_bilayer}, see Fig.\ \ref{fig-cond}a. 

On the light of the above discussion, we proposed in Ref.\
\onlinecite{benfatto_kt_bilayer} that a more reasonable explanation for the
rounding effects in the superfluid density and the large transient
region in $\sigma_1(\o)$ observed in the experiments is the sample inhomogeneity. In the case
of cuprate systems such inhomogeneity is suggested by
tunneling measurements in other families of cuprates,\cite{yazdani_07}
where  Gaussian-like fluctuations of the local gap value are
observed. Recently similar local-gap distributions have been reported also
in tunneling experiments in thin films of strongly-disordered conventional
superconductors\cite{sacepe_stm_prl08,sacepe_stm_natcom10,sacepe_stm_natphys11,pratap_prl11,pratap_stm_cm11},
showing that an intrinsic tendency towards mesoscopic 
inhomogeneity appears even for systems with homogeneous intrinsic disorder. 
Even though the issue of the microscopic origin of this effect is
beyond the scope of our work, nonetheless we found that a
Gaussian-like distribution of the superfluid-stiffness $J_0$ values around a given $\bar
J$ can account very well for the data of
Ref.~[\onlinecite{lemberger_bilayer}] on two-unit-cell thick films of
YBa$_{2}$Cu$_{3}$O$_{6+x}$. Thus, we compared with the experiments
the quantity $J_{inh}(T)=\int dJ_0 P(J_0)J(T,J_0)$, where each $J(T,J_0)$
curve is obtained from the 2D RG equations using a
bare superfluid stiffness $J=J_0-\a T^2$. Each initial value $J_0$ has a
probability $P(J_0)=\exp[-(J_0-\bar J_0)^2/2\s^2]/(\sqrt{2\pi} \s)$ of being
realized, where the bare average stiffness follows the typical temperature
dependence due to quasiparticle excitations in a (disordered) $d$-wave
superconductor, i.e. $\bar J(T)=\bar J_0-\a T^2$, with
$\bar J_0$ and $\a$ fixed by the experimental data at low $T$, where
$J_{exp}(T)$ is practically the same as $\bar J(T)$. Using a
variance $\s=0.05 \bar J_0$ we obtained a very good agreement with the
experiments near the transition, as far as both both the tail of $\l^{-2}$
and the position and width of $\s_1(\omega)$ are concerned, see Fig.\
\ref{fig-cond}b.

\begin{figure}[ht]
\begin{center}
\parbox{2.1in}{\includegraphics[width=6cm]{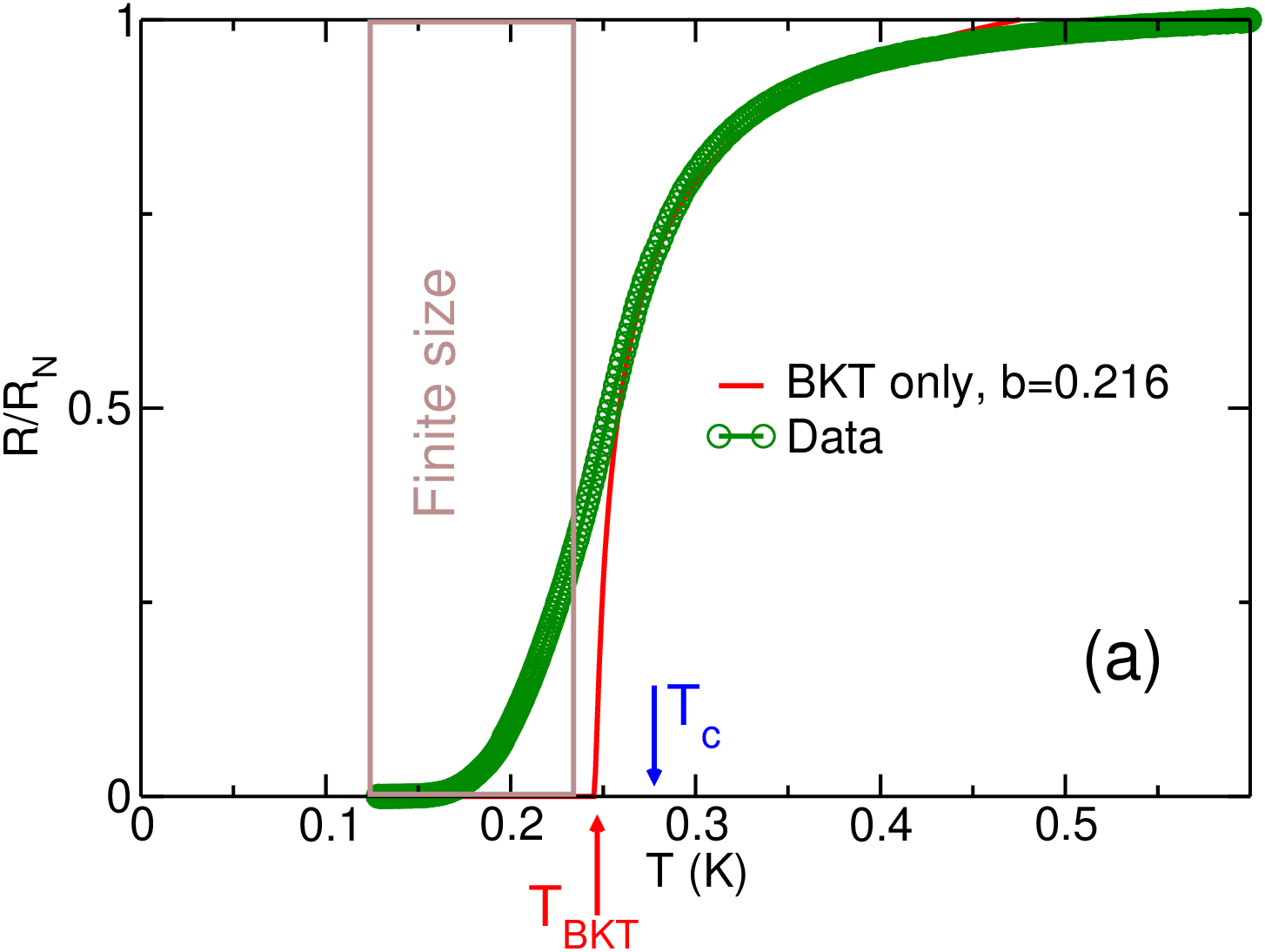}}
\hspace*{30pt}
\parbox{2.1in}{\includegraphics[width=6cm]{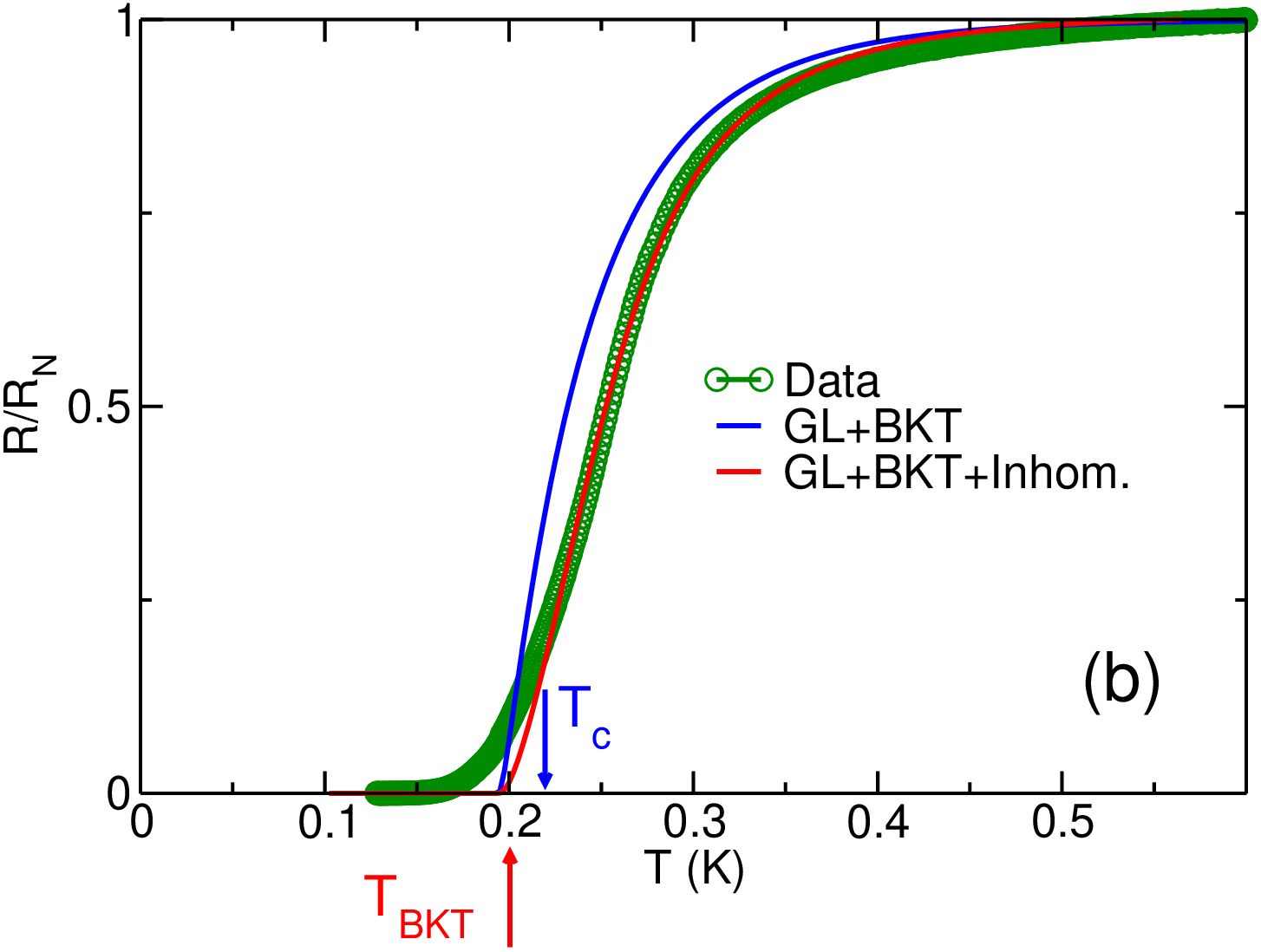}}
\end{center}
\caption{Analysis of the paraconductivity effect on the resistivity of
  SC interfaces at the
  $T_{BKT}$ transition. The experimental data for the resistivity $R$ normalized to
  the normal-state value $R_N$ are taken from Ref.\
  [\onlinecite{triscone_nature08}]. Panel (a) shows an example of the approach proposed in
  [\onlinecite{triscone_science07,triscone_nature08,schneider_finitesize09}],
  where the whole range of temperatures above $T_{BKT}$ is dominated by SC
  fluctuations having BKT character, and finite-size effects are
  responsible for the resistive tail below $T_{BKT}$. However, the $b$
  value obtained by the BKT fit based on Eq.\ \pref{xibkt} for the
  correlation length implies, according to Eq.\ \pref{btheo}, the $T_c$
  value marked by the blue arrow, invalidating thus the assumption itself
  that the whole fluctuation regime has BKT character. Panel (b) is
  taken from Ref.\ \onlinecite{caprara_prb11} (Fig. 5.1 in that paper) and elucidates the 
approach developed in Refs.\
  \onlinecite{benfatto_kt_interfaces,caprara_prb11}, based on the interpolation
  scheme \pref{xihn} between BKT and GL fluctuation. The curve labeled
``GL+BKT+inhom'' has been computed by solving a
  random-resistor network problem with a Gaussian distribution
  ($\sigma=0.035$ K) of the
  critical temperatures centered around
$\bar T_{BKT}=0.2$ K, marked by an arrow along with the corresponding
$\bar T_{c}=0.22$ K. The paraconductivity of each resistor follows
Eq.\ \pref{xihn} with $b=0.25$, which 
  according to Eq.\ \pref{btheo} corresponds to $t_c=0.1$ when
$\mu/J_s\simeq 2$, an intermediate value between the $XY$ model
  \pref{defmu} and BCS \pref{mu} estimates. The curve labeled
  ``GL+BKT'' corresponds to the homogeneous case with a transition
  temperature $\bar T_{BKT}$.}
\label{fig-scheme}
\end{figure}

Even though our approach to the issue of inhomogeneity is quite
phenomenological, it has been shown more recently that it accounts very
well also for the superfluid-density behavior in films on
NbN\cite{pratap_apl10,pratap_bkt_prl11}. In all these cases indeed the rounding
effect due to inhomogeneity is far more relevant than any finite-size
effect due to screening or finite-frequency probes, explaining the
lack of a very sharp jump even for relatively weakly-disordered films
(see Fig.\ \ref{fig_mu}(b) above). 
Quite interestingly a similar physical picture seems to be relevant also
for a completely different class of materials, i.e. the SC
metal-oxides interfaces, where experimental data for the superfluid density
are not yet available, but the nature of the SC transition can be
investigated in an indirect way by means of the analysis of the paraconductivity
above $T_{BKT}$\cite{triscone_science07}. For these systems as well a BKT transition is likely to be
expected, since the SC interfaces are very thin\cite{triscone_science07}, of the order of 15 nm,
which is much lower than the SC coherence length $\xi_0\simeq 70$
nm. Moreover, despite what occurs in thin films of conventional
superconductors\cite{pratap_bkt_prl11}, the drop of the resistivity above the transition is very
smooth, suggesting that inhomogeneities on a mesoscopic scale broaden
considerably the transition, as we discussed in details in Ref.\
[\onlinecite{benfatto_kt_interfaces}]. On this respect, it is worth recalling that
any analysis of the BKT transition based on the paraconductivity data alone
can suffer of the unavoidable lack of knowledge about the exact extension
of the BKT fluctuation regime. Indeed, as we mentioned in the introduction,
the contribution of SC fluctuations to the paraconductivity is encoded in
the temperature dependence of the correlation length $\xi(T)$ both within
the BKT and GL theory, see Eq.\ \pref{scf}. Thus, if the transition has BKT
character, one should expect a crossover from the BKT exponential temperature
dependence of $\xi(T)$,
\be
\lb{xibkt}
\xi_{BKT}=A\exp(b/\sqrt{t}), \quad t\equiv \frac{T-T_{BKT}}{T_{BKT}},
\ee
to the power-law GL one, $\xi_{GL}\sim \xi_0 T_c/(T-T_c)$, where $T_c$ is
the BCS mean-field critical temperature. In the case of thin films of
superconductors the BKT regime is in practice restricted to a very small range
of temperatures near $T_{BKT}$, due to the fact that $T_c$ at most
twenty per cent larger than $T_{BKT}$. In the case of the SC
interfaces it has been proposed instead  in the recent
literature\cite{triscone_science07,triscone_nature08,schneider_finitesize09}
that $T_c$ is far larger than $T_{BKT}$, so that the {\em whole}
fluctuation regime above $T_{BKT}$ is dominated by BKT fluctuations alone,
while the large tails observed experimentally near $T_{BKT}$ should be
ascribed to finite-size effects, see Fig.\ \ref{fig-scheme}a. There is
however a serious drawback in such interpretation, that originates from an
incorrect application of the BKT relation \pref{xibkt} in the analysis of
the experimental data. As we showed in details in Ref.\
[\onlinecite{benfatto_kt_interfaces}] by means of a RG analysis of the
correlation length, the parameter $b$ which appears in the
usual BKT formula \pref{xibkt} is directly connected to the distance
$t_c=(T_c-T_{BKT})/T_{BKT}$ between the mean-field $T_c$ and the BKT
transition temperature and to the value of the vortex-core energy with
respect to the superfluid stiffness:
\be
\lb{btheo}
b_{theo}\sim \frac{4}{\pi^2}\frac{\mu}{J_s}\sqrt{t_c}, \quad t_c=\frac{T_c-T_{BKT}}{T_{BKT}}.
\ee
As a consequence, $b$ increases when  the BKT fluctuation regime
extends in temperature. In the typical fits of the resistivity proposed in Ref.\
\cite{triscone_nature08,schneider_finitesize09} one obtains values of the
$b$ parameter that, according to Eq.\ \pref{btheo} above, would imply a
mean-field $T_c$ very near to $T_{BKT}$, see for instance Fig.\
\ref{fig-scheme}a. In other words, the fit  leads to values of $b$ that
contradict the a-priori assumption that the whole fluctuations regime is
dominated by BKT fluctuation, as described by Eq.\ \pref{xibkt}. Moreover,
also the interpretation of the tails extending below $T_{BKT}$ as due to
finite-size effects\cite{schneider_finitesize09} 
leads to unphysical low values for the size of the
homogeneous domains.\cite{benfatto_kt_interfaces}  
One would have thus to invoque a BKT transition of a completely
differente nature (dislocations of a vortex crystal\cite{gabay_melting}), as it has been
suggested in Ref.\ [\onlinecite{triscone_science07}]. However, even in this
case there is no theoretical understanding how to reconcile such
contradictory numbers obtained in the analysis of the resistive
transition. 

An alternative, and less far fetched, 
interpretation of the resistivity data can be based instead on the use of
the interpolation formula for $\xi(T)$ proposed 
long ago by Halperin and Nelson\cite{halperin_ktfilms}
\be
\lb{xihn}
\frac{R}{R_N}=\frac{1}{1+(\xi/\xi_0)^2}, \quad 
\xi(T)=\xi_0 A \sinh \frac{b}{\sqrt t}.
\ee
One can easily see that
Eq.\ \pref{xihn} reduces to $\xi_{BKT}$ for small $t$ and to $\xi_{GL}$
for large $t$.  Using the estimate \pref{btheo} for the $b$ parameter one
realizes that the crossover occurs around $t\simeq b^2 \propto t_c$,
so that for (realistic) small values of $t_c$ most of the fluctuation regime
is dominated by GL-type fluctuations. Moreover, we attributed the
broadening of the transition to the effect of inhomogeneity, that induces a
distribution of possible realizations of local $R/R_N$ values
corresponding to the local $J$ values discussed above. An example is shown
in Fig.\ \ref{fig-scheme}b, where the average resistivity has been computed
by solving a random-resistor-network model for a set of resistors
undergoing a metal-superconductor transition, as discussed recently in
Ref.\ [\onlinecite{caprara_prb11}]. It must be emphasized that the issue of the physical value of the $b$
parameter in the BKT expression \pref{xibkt} of the correlation length has been
often overlooked in the literature also in contexts of other
materials\cite{triscone_fieldeffect,iwasa_natmat10}, leading
to questionable conclusions concerning the existence of a BKT physics based only
on erroneous fits of the resistivity. In contrast, a recent analysis
on NbN films\cite{pratap_bkt_prl11}, based on the comparison between the estimate of $b$ deduced
from the analysis of the superfluid density below $T_{BKT}$ and the
resistivity above $T_{BKT}$  has confirmed the theoretical estimate
\pref{btheo} of the $b$ parameter, that must be used as an unavoidable
constraint in any BKT fit of the paraconductivity. 

\section{The case of a finite magnetic field}
\label{magfield}

As we discussed in the previous Sections, the sine-Gordon model provided us
with the most appropriate formalism to investigate not only the role of a
vortex-core energy value different from the $XY$-model one, but also the
occurrence of BKT physics in the presence of a relevant perturbation (in
the RG sense), as the interlayer coupling. It is then worth asking the
question of the possible relevance of such a mapping for an other typical
relevant perturbation, i.e. a finite external magnetic field. The nature of
the BKT physics in the presence of a magnetic field has been of course
already investigated in the
past\cite{review_minnaghen,doniach_films,minnaghen_finitefield}, with a
renewed interest in the more recent literature\cite{sondhi_kt,huse_prb08,podolsky_prl07} triggered
once more by experiments in cuprate superconductors carried out at finite
field, as the measurement of the Nerst effect or the magnetization.
Unfortunately, contrary to the case of the $\bB=0$ transition, the efforts
have been partly unsatisfactory. In particular most of the literature on
the subject rested on the use of the mapping into the Coulomb-gas problem,
where the effects of the magnetic field can be incorporated as an excess of
positive charges, in analogy with the finite population of vortices with a
given vorticity due to the presence of the external field. There is however
a fundamental drawback of this approach, since it gives the physical
observables as a function of the magnetic induction $\bB$ instead of the
magnetic field $\bH$. This is of course not convenient at low applied
field, since inside the superconductor $\bB$ vanishes even for a finite external field
$\bH$.  Motivated by this observation and by the occurrence of an anomalous
non-linear regime for the magnetization measured in cuprate systems, we
showed recently\cite{benfatto_kt_magnetic} that a suitable mapping into the
sine-Gordon model provides a very simple and physically transparent way to
deal with the finite magnetic-field case, leading to a straightforward
definition of the physical observables and clarifying the role of both $\bB$
and $\bH$. Since the construction of the mapping has not been shown in
Ref.\ [\onlinecite{benfatto_kt_magnetic}] due to space limitation, we shall discuss
here both the model derivation and its basic physical consequences. This
will complete our overview of the sine-Gordon approach to the BKT physics,
and it will allow us also to clarify some aspects of the charged superfluid
that have not been underlined yet in the previous Sections. 

As a starting point we use again the $XY$ model \pref{xy}, where the coupling to $B$ is introduced via the
minimal-coupling prescription for the vector potential $\bA$, 
\be
\lb{hamf}
H=J\sum_{<i,j>} [1-\cos(\th_i-\th_j-F_{ij})]
\ee
where 
\be
\lb{defF}
F_{ij}=F_\mu(r)=\frac{2\pi}{\Phi_0}\int_r^{r+\mu}\bA\cdot d\bl\approx
\frac{2\pi a}{\Phi_0}\bA_\mu(r).
\ee
Here $\mu=\hat x,\hat y$ is the vector from one site to the neighboring one
and $\Phi_0=hc/2e$ is the flux quantum. As a first step we make use of the
Villain approximation, that amounts to replace the cosine in Eq.\
\pref{hamf} with a function having the same minima for each 
multiple $2\pi m$ of the gauge-invariant phase difference, i.e.
\be 
\exp [-\beta J
(1-\cos(\theta_i-\theta_j-F_{ij})]\Ra \sum_{m=-\infty}^{m=\infty} \exp
\left[-\frac{\beta J}{2} (\theta_i-\theta_j-F_{ij}-2\pi m)^2 \right].
\ee
By following the analogous derivation given in Ref.\ [\onlinecite{nagaosa_book}]
for the case $\bA=0$, we make use of the Poisson summation formula 
\be
\lb{poiss}
\sum_{m=-\infty}^{m=\infty}  h(m)=\sum_{l=-\infty}^{l=\infty} \int dz h(z)
e^{2\pi il z},
\ee
and by performing explicitly the integration over $z$  for each link $ij$ we
are left with the following structure of the partition function:
\be
\lb{part}
Z=Z_J\sum_{\{l_{ij}\}}\int_0^{2\pi} d\th_1 ... d\th_N
e^{-\sum_{<i,j>}\frac{l_{ij}^2}{2\beta J}+i\sum_{\langle i,j \rangle}l_{ij}
(\th_i-\th_j-F_{ij})},
\ee
where $N$ is the number of lattice sites and $l_{ij}$ are $2N$ integer
variables defined for each link $ij$. The prefactor $Z_J=[(1/\beta
J)^{1/2}]^{2N}=(1/\beta J)^N$ accounts for the $z$ integration above
for each link. Since each $\theta$ variable is
defined on a period, we have that $\int_0^{2\pi} d\theta
e^{i\theta \alpha}=\delta(\alpha)$. As a consequence,  the integration over the
$\theta_i$ variables in the above equation leads to $N$ constraint equations:
\be
\lb{const}
\sum_\mu l_\mu(\br)-l_\mu(\br-\mu)=0,
\ee
which are the discrete equivalent of $\nb\cdot \bl=0$. These equations can
be satisfied by defining for each site of the lattice a field $n$
such that:
\bea
\lb{lx}
l_x(\br)=n(\br)-n(\br-y),\\
\lb{ly}
l_y(\br)=n(\br-x)-n(\br),
\eea
i.e. the discrete equivalent of the relation $\bl=\nb\times (n \hat z)
$. By means of Eqs. \pref{lx}-\pref{ly} Eq.\ \pref{part} can be then rewritten
as:
\be
Z=Z_J\sum_{\{n(\br)\}}e^{\sum_{\br,\mu}\frac{1}{2\beta J}(\D_\mu n(\br))^2}
e^{-i\frac{2\pi a}{\Phi_0}\sum_\br \hat z\cdot (\bA\times \D n)},
\ee
where $\D_\mu=n(\br+\mu)-n(\br)$ is the discrete derivative. Finally, we can use again
the Poisson summation formula to get:
\be
\lb{partphi}
Z=Z_J\int\DD\phi(\br)\sum_{\{m(\br)\}}
e^{-\sum_{\br,\mu}\frac{1}{2\beta J}(\D_\mu \phi(\br))^2
+2\pi i\sum_\br \phi(\br)m(\br)
-i\frac{2\pi a}{\Phi_0}\sum_\br \hat z\cdot (\bA\times \D \phi)},
\ee
where the $m(\br)$ variables assume arbitrary positive and negative
integer values.
Before completing the mapping into the sine-Gordon model we notice that in
the above equation the $\phi$ variables can be integrated out exactly. By
rewriting the last term in the action as $\sum_\br \hat z\cdot (\bA\times
\D \phi)= \sum_\br \phi \hat z \cdot (\D \times \bA) =B a \sum_\br \phi$ one
can easily see that the final result is given by
\be
\lb{partcg}
Z=Z_\pll\sum_{\{m(\br)\}}e^{-2\pi^2\beta J\sum_{\br,\br'}
[m(\br)-f]U(r-r')[m(\br')-f]}, \quad f=\frac{Ba^2}{\Phi_0}
\ee
Here $Z_\pll=Z_J\Pi_{q>0} (\beta J/\bq^2)\equiv \Pi_{q>0} (1/\beta J
\bq^2)$ is the overall contribution due to the longitudinal
excitations. In full analogy with the result of Sec.\ \ref{mapping},
the longitudinal modes decouple from the transverse ones, so their
contribution $Z_\pll$ to the partition function can be discarded in
what follows.  The function $V(\br)=\sum_\bk e^{i\bk \cdot\br} U(\bk)$
is defined trough the Fourier transform $U^{-1}(\bk)=(4-2\cos
k_x-2\cos k_y)$ of the $\Delta_\mu$ operator on the square lattice,
i.e.
\be
\lb{ur}
V(\br)=\int \frac{d^2\bk}{(2\pi)^2}\frac{e^{i\bk\cdot \br}}{[4-2\cos
    k_x-2\cos k_y]}.
\ee
Eq.\ \pref{partcg} generalizes the Coulomb-gas formula \pref{coulomb} above
to the case of a finite magnetic field. As we did in Sec. \pref{mapping}, we can separate in
$V(\br)$ the singular part in $\br=0$ by defining the regular function
$G(\br)$, i.e. $V(\br)=V(0)+G(\br)$. Then one sees that 
\bea
& &\sum_{\br,\br'}[m(\br)-f]V(r-r')[m(\br')-f]=\nn\\
&=&V(0)\left[\sum_\br (m(\br)-f)\right]^2+
\sum_{\br,\br'}[m(\br)-f]G(r-r')[m(\br')-f]\nn
\eea
It then follows that also in this case only neutral configurations have a statistical weight
different from zero. However, in the presence of a magnetic field the neutrality condition reads:
\be
\lb{neutr}
\sum_\br (m(\br)-f)=N_v-\frac{Ba^2N}{\phi_0}=0
\ee
which means that the total flux $N_v\Phi_0$ carried out by the (unbalanced) vortices
equals the total flux $Ba^2 N$ of the magnetic field
across the sample. The definition \pref{ur} allows us also to determine the
value \pref{defmu} of the chemical potential $\mu$ in the lattice $XY$ model. Indeed, one can see that at
the scale of the lattice spacing $V(\br)-V(0)$ gives:
\bea
V(\br=\hat x)-V(0)&=&
\int \frac{d^2\bk}{(2\pi)^2}\frac{\cos k_x-1}{[4-2\cos
    k_x-2\cos k_y]}=\nn\\
\lb{vlattice}
&=&\frac{1}{2}\int \frac{d^2\bk}{(2\pi)^2}\frac{\cos k_x+\cos k_y-2}{[4-2\cos
    k_x-2\cos k_y]}=-\frac{1}{4},
\eea
so that from Eq. \pref{partcg} it follows that the cost to put two vortices
at distance $a$ apart is $\beta\mu=\beta\pi^2 J/2$, consistent with the
value \pref{defmu} that we quoted above. 

Let us now go back to the Eq.\ \pref{partphi} and let us complete the
mapping into the sine-Gordon model. We notice that in Eq.\ \pref{partphi}
the   variable $\phi$ is still defined
on the square lattice. We can however resort to a continuum approximation
by taking into account the energetic cost $\mu$ of the vortex creation on
the shortest length scale of the problem via a chemical-potential like term
$e^{\ln y\sum_\br m^2(\br)}$ in Eq.\ \pref{partphi}:
\be
\lb{partphi2}
Z=\int\DD\phi(\br)\sum_{\{m(\br)\}}
e^{-\sum_{\br,\mu}\frac{1}{2\beta J}(\D_\mu \phi(\br))^2
+2\pi i\sum_\br \phi(\br)m(\br)
-i\frac{2\pi a}{\Phi_0}\sum_\br \hat z\cdot (\bA\times \D \phi)}
e^{\ln y\sum_\br m^2(\br)}
\ee
The vortex-core energy term favors the formation of vortices of smallest
vorticity, i.e. $m(\br)=0,\pm 1$. If we limit ourselves to this case the sum over the integer
variables $m(\br)$ can be performed explicitly as:
\be
\sum_{m(\br)=0,\pm 1}e^{\ln ym^2+2\pi i m \phi}=1+2y\cos(2\pi\phi)\approx
e^{2y\cos(2\pi\phi)} 
\ee
Inserting this into Eq.\ \pref{partphi}, taking the limit of the
continuum $(\sum_\br\ra (1/a^2)\int  d^2\br)$, and rescaling $\phi\ra
\phi/\pi$  we finally obtain a partition function expressed in terms of the
$\phi$ field only, that generalizes Eq.\ \pref{sg} above: 
\be
\lb{sb}
S_B=\int d\br dz 
\left[\frac{(\nb \phi)^2}{2\pi K} \!-\!\frac{g}{\pi a^2}\cos 2\phi
+\frac{2 i}{\Phi_0}\bA\!\cdot\!(\nb \times   \hat z\phi)
\right] \delta(z),\,
\ee
where we used again the definitions \pref{defk}-\pref{defg} for $K,g$.
In Eq.\ \pref{sb} we added also the explicit $z$ dependence of the action,
which is needed since the $\bA$ field depends in general
also on the $z$ out-of-plane coordinate. The $\delta(z)$ function
gives the proper boundary conditions for a truly 2D case (where there is no
SC current outside the plane), while in the physical case of a SC film of finite
thickness $d$ we shall assume that the sample quantities are averaged over
$|z|<d/2$. 

The action \pref{sb} and its corresponding partition function  $Z=\int \DD\phi
e^{-S_B}$ allow for a straightforward definition of the
physical observables at finite field. For example, the electrical
current follows as usual from the functional derivative of $F=-(1/\beta)\ln
Z$ with respect to
the gauge field $\bA$, i.e
\be
\lb{js}
\bJ_s(\br,z)=-c\frac{\pd F}{\pd \bA(\br,z)}=-
\frac{2ick_BT}{\Phi_0}\langle \nb\times \hat z\phi(\br)\rangle\delta(z).
\ee
The Eq.\ \pref{js} makes it evident that the current is purely
transverse, as expected for vortex excitations, according to the
discussion given in Sec. \ref{mapping} above. A second quantity that
can be easily obtained is the magnetization $\bM=(\bB-\bH)/4\pi$, defined as the
functional derivative of $F$ with respect to $\bB(\br,z)=\nb \times
\bA$. By integrating by part and using the identity
$\int (\bA\times\nb\phi)\cdot \hat z=\int \nb\phi\cdot (\hat z\times \bA)=
-\int \phi\nb\cdot (\hat z\times \bA)=\int \phi \hat z\cdot
(\nb\times\bA)$, we can rewrite the last term of Eq.\ \pref{sb} 
as
\be
\lb{sbb}
S_B=\int d\br dz 
\left[\frac{(\nb \phi)^2}{2\pi K} \!-\!\frac{g}{\pi a^2}\cos 2\phi
+\frac{2 i}{\Phi_0} \bB(\br,z)\cdot \hat z \phi(\br)
\right] \delta(z),\,
\ee
As a consequence the functional derivative with respect to $\bB$ gives
immediately 
\be
\lb{mag2}
\bM(\br)=-\frac{1}{d}\int dz \frac{\pd F}{\pd \bB(\br,z)}=-\hat z
\frac{2ik_BT}{d\Phi_0}\left\langle\phi(\br)\right\rangle,
\ee
and analogously the uniform magnetic susceptibility $\chi=\pd M_z/\pd
B_z$ is:
\be  
\lb{chim}
\chi=-(4k_B T)/(d\phi_0^2)\int d\br \left[\langle
\phi(\br)\phi(0)\rangle -\langle \phi(\br)\rangle \langle \phi(0)\rangle
\right].
\ee
Notice that the total magnetic moment $\cal M$ associated to the current $\bJ_s$ is
${\cal M}= (1/2c)\int d\br dz \, (\br\times \bJ_s)$\cite{landau_iv}. Using the
definition \pref{js} of $\bJ_s$, one can easily verify that ${\cal M}=\int
d^2\br dz \bM$, i.e.  the magnetization is the density of magnetic moment
of the sample, as expected\cite{landau_iv}. 
Finally, by exploiting the fact that $e^{-\beta\mu}e^{\pm i2\phi}$ is the operator
which creates up and down vortices with density $n_\pm$ respectively, we
have a straightforward definition of the average vortex number
$n_F=a^2(\langle n_+\rangle +\langle n_-\rangle)$ and of the excess vortex
number $n=a^2(\langle n_+\rangle -\langle n_-\rangle)$ per unit cell as a
function of $\phi$ as:
\be
\lb{dens}
n_F=2 e^{-\beta\mu} \langle \cos (2\phi)\rangle, \quad
n=2e^{-\beta\mu} \langle \sin (2\phi)\rangle.
\ee

In Eq.\ \pref{mag2} above
 the average value of $\phi$ is computed with the action
\pref{sb}, so that the magnetization is given 
as a function of the magnetic induction
$\bB$. However, as we mentioned above, at low external field it would
be more convenient to compute $\bM$ as a function of the applied field
$\bH$. This can be achieved by using 
the Gibbs free energy $\GG=-k_BT\ln Z_G$, where the partition function
$Z_G$ includes also the contribution of the electromagnetic field $Z_G=\int \DD\phi\DD Ae^{-S}$
and:
\be
\lb{saf}
S=S_B+\int{d\br dz}\left\{
\frac{(\nb\times\bA)^2}{8\pi k_BT}
-\frac{(\nb\times \bA)\cdot\bH}{4\pi k_BT}\right\}.
\ee
Before discussing explicitly the case of a finite external field $\bH$
we would like to stress how Eq.\ \pref{saf} gives also a very
convenient description of the role of charged supercurrents in a 2D
superconductor.
Indeed, even when $\bH=0$ the electromagnetic field $\bA$ in Eq.\
\pref{saf} above describes the magnetic field created by the current
themselves in a charged superfluid.  In this case, since the SC
currents live in the plane, also $\bA$ is a two-dimensional vector, so
that $\nb\times\bA=(-\pd_z A_y,\pd_z A_x,\pd_x A_y-\pd_y
A_x)$. Moreover, is we choose the Coulomb (or radial) gauge
$\nb\cdot\bA=0$, we have that in Fourier space $(\bk_\pll\times
\bA)^2=k_\pll^2 \bA^2$.  We can then rewrite in Fourier space the
terms in $\bA$ of Eq.\ \pref{saf} at $\bH=0$ as:
\be
\int\frac{d^3\bk}{(2\pi)^3}\left\{
- \frac{2}{\Phi_0} \phi(\bk_\pll)|\bk_\pll\times A(\bk_\pll,k_z)|-
\frac{(k^2_z+\bk_\pll^2)}{8\pi T}\bA^2(\bk_\pll,k_z)\right\}.
\ee
By integrating out $\bA$ at Gaussian level we obtain a $\phi^2$
contribution to the action of the form:
\be
\lb{gausscontr}
\int\frac{d^3\bk}{(2\pi)^3}
\frac{8\pi T}{\Phi_0^2}\frac{\bk_\pll^2}{(k^2_z+\bk_\pll^2)}|\phi(\bk_\pll)|^2.
\ee
Since $\phi$ depends on $\bk_\pll$ only, we can integrate out $k_z$ and
obtain that the overall Gaussian action for the $\phi$ field reads:
\be
\lb{screen}
S_G=\int\frac{d^2\bk_\pll}{(2\pi)^2}\frac{1}{2\pi
  K}[k_\pll^2+k_\pll\L^{-1}]|\phi(\bk_\pll)|^2,
\ee
where we defined:
\be
\lb{defl}
\frac{1}{\L}=\frac{8\pi^2 KT}{\Phi_0^2}\equiv\frac{d}{2\lambda^2}.
\ee
The last equality follows from the definitions \pref{defk} and
\pref{defj} of $K=\pi J/T$ and $J=\Phi_0^2d/16\pi^3\l^2$,
respectively, and it allows us to identify $\L$ with the so-called
Pearl screening length\cite{pearl}. Indeed, due to the $k_\pll\L^{-1}$
terms in Eq.\ \pref{screen}, one sees that the potential between
vortices, which according to the above discussion is the Fourier
transform of the Gaussian $\phi$ propagator, decays as $e^{-r/\L}$ at
scales $r\gg\L$, instead of the usual $\log r$ dependence observed at
all length scales in neutral superfluids. 

Let us go back now to the case of a finite external field in Eq.\
\pref{saf} and let us integrate again the gauge field $\bA$. Since now
it is present an additional term $-(i/4\pi T) \bA(\bk) \cdot
(\bk\times \bH(-\bk))$ in $S$, one obtains  the following contribution to
the action:
\be
\frac{8\pi T}{k^2}\left|\frac{\phi(-\bk_\pll)}{\phi_0}(\hat z\times \bk_\pll)
+\frac{i}{8\pi T}(\bk\times \bH(-\bk))\right|^2.
\ee
The quadratic term in $\phi$ corresponds to Eq.\ \pref{gausscontr}
above, leading to the screening of the vortex potential. 
The remaining terms can be written as:
\be
\lb{sph}
S_{\phi-H}=\int\frac{d^3\bk}{(2\pi)^3}
\frac{2 i}{\Phi_0 k^2}(\hat z\times \bk_\pll \phi(\bk_\pll))\cdot
(\bk\times \bH(-\bk)),
\ee
and
\be
\lb{shh}
S_{H-H}=\int\frac{d^3\bk}{(2\pi)^3}
\frac{1}{8\pi T}\frac{(\bk\times \bH)^2}{k^2}.
\ee
Using the identity
\be
\int\frac{d^3\bk}{(2\pi)^3} F_1(\bk)F_2(-\bk)\frac{1}{\bk^2}=
\int d^3\br d^3\br' F_1(\br)F_2(\br')\frac{1}{4\pi |\br-\br'|},
\ee
and the Maxwell equation relating the magnetic field $\bH$ to the 
distribution of the external current $\bJ_{ext}$ producing the field itself
\be
\lb{max}
\nb \times\bH=\frac{4\pi}{c} \bJ_{ext},
\ee
one can easily see that Eq.s \pref{sph} and Eq.\ \pref{shh} can be written in real space as:
\be
\lb{sreal}
S_{\phi-H}=2i\int d^3\br  d^3\br' \frac{[\hat z\times \nb \phi(\br)]\cdot
[\nb'\times \bH(\br')]}{4\pi |\br-\br'|},
\ee
and
\be
\lb{sh2}
S_{H-H}=-\frac{1}{8\pi T}
\int d^3\br  d^3\br' \frac{\bJ_{ext}(\br)\cdot \bJ_{ext}(\br')}
{4\pi |\br-\br'|} ,
\ee
respectively. By integration by part Eq.\ \pref{sreal} can be rewritten as:
\bea
S_{\phi-H}&=&\frac{2i}{c}
\int d^3\br  d^3\br' \frac{[\hat z\times \nb \phi(\br)]\cdot
\bJ_{ext}(\br')}{|\br-\br'|}=\nn\\
&=&-\frac{2i}{c}
\int d^3\br  d^3\br' \phi(\br)\nb_\br\cdot \frac{[\bJ_{ext}(\br') \times 
\hat z]}{|\br-\br'|}=\nn\\
&=&\frac{2i}{c}
\int d^3\br  d^3\br' \phi(\br)\hat z \cdot \frac{[\bJ_{ext}(\br') \times 
(\br-\br')]}{ |\br-\br'|^3}=\nn\\
\lb{sph2}
&=&\frac{2i}{\Phi_0}\int d^3 \br \phi(\br) \hat z\cdot \bH^0(\br).
\eea
In the last equality of Eq.\ \pref{sph2} we introduced the reference
field ${\bf H^0}$, which corresponds to the magnetic field generated
by the same distribution of currents $\bJ_{ext}$ {\em in the vacuum}.
According to the Laplace formula, $\bH^0$ is given exactly by
\be
\lb{h0}
\bH^0(\br)=\frac{1}{c}\int
d^3\br' \frac{\bJ_{ext}(\br') \times 
(\br-\br')}{ |\br-\br'|^3},
\ee
leading to Eq.\ \pref{sph2} above. One can also recognize in the term \pref{sh2} the
magnetic energy density associated to the reference field
$\bH^0$,
\be
\lb{enh0}
S_{H-H}=-\frac{1}{8\pi T}\int d^3\br (\bH^0)^2.
\ee
Indeed, since $\bH^0$ is the field created by the currents
$\bJ_{ext}$ in the vacuum it satisfies $\nb \cdot \bH^0=0$, so that
 $(\bk\times \bH^0)^2=(\bH^0)^2 k^2$. Thus $\bH^0(\bk)^2=(\nb \times
 \bH^0)/k^2 = \bJ_{ext}(\bk)^2/k^2$ that is the Fourier transform of
 Eq.\ \pref{sh2}. In summary, the full sine-Gordon action after integration of the gauge
field can be rewritten as:
\bea
S=\int\frac{d^2\bk_\pll}{(2\pi)^2} \frac{k_\pll^2+k\L^{-1}}{2\pi K}|\phi(\bk_\pll)|^2
-\frac{g}{\pi a^2}\int d\br \cos 2\phi\nn\\
\lb{sfin}
+\frac{2 i }{\Phi_0} \int d\br\, \phi \,\hat z\cdot \bH^0(\br,z=0)
-\int d\br dz \frac{(\bH^0)^2}{8\pi k_B T}.
\eea
Eq.\ \pref{sfin} is the desired result to be used to evaluate the
physical observable as a function of the reference field $\bH^0$.
Once more, the sine-Gordon mapping turns out to provide a quite
powerful framework for the investigation of the BKT physics of a
superconductor embedded in an external field. Indeed, apart from the
fact that it includes automatically the screening effect of the
supercurrents discussed above, the action \pref{sfin} expressed in
terms of $\bH^0$ has two main advantages. First of all, $\bH^0$ is the
field quoted in the experimental measurements, since what is known
{\em a priori} are only the generating currents $\bJ_{ext}$. Indeed,
$\bH^0$ does not coincide in general with the real field $\bH$ even
outside the sample, since the $\bH$ configuration takes into account
also the field exclusion from the SC sample, the so-called
demagnetization effects. For simple sample geometries one can include
these effects in a demagnetization coefficient $\eta$, and write in 
general the following relation between $\bB=\bH+4\pi \bM$ and $\bH^0$
\cite{landau_iv}:
\be
\lb{gen}
(1-\eta)\bH+\eta\bB=\bH^0 \Ra \bB=\bH^0+4\pi(1-\eta)\bM.
\ee
In the complete
Meissner phase one has $\bB=0$, which implies $-4\pi\bM=\bH$. However, from
Eq.\ \pref{gen} it follows that $\bH=\bH^0/(1-\eta)$ so that:
\be
\lb{meiss}
\bM=-\frac{1}{4\pi}\frac{\bH^0}{1-\eta}.
\ee
While for a cylinder
$\eta=0$ and $\bH=\bH^0$, for a film of thickness $d$ and transversal
dimension $R$ one has that $\eta\sim 1-d/R$, so one
expects to find $\bM\sim (R/d) \bH^0$ below $H_{c1}$, i.e. a much
smaller critical field with respect to the same system in the 3D
geometry\cite{fetter_films,landau_iv}. 
Since the magnetization
$M$ calculated from Eq.\ \pref{sfin} is already a function of $\bH^0$,
it will include automatically all the demagnetization effects and the
complications of the thin-film geometry. 

These properties have been derived in Ref.\
[\onlinecite{benfatto_kt_magnetic}], where the magnetization has been computed
by means of a variational approximation for the cosine term in the
model \pref{sfin}.  While we refer the reader to Ref.\
[\onlinecite{benfatto_kt_magnetic}] for more details concerning these
calculations, we would like to mention here one particular result,
that is related to the discussion of the previous Sections. It
concerns the behavior of the field-induced magnetization above
$T_{BKT}$, that is expected\cite{halperin_ktfilms} to be proportional to $H$ with a
coefficient depending on the SC correlation length:
\be
\lb{hlin}
M=-\frac{k_B T}{d\Phi_0^2}\xi^2 H. 
\ee
In full analogy with the paraconductivity discussed in the previous
Sections, the functional dependence of the low-field magnetization $M$
on the BKT correlation length $\xi$ in Eq.\ \pref{hlin} is the same as
in the GL theory. While this result was already known in the
literature\cite{halperin_ktfilms,sondhi_kt}, our calculations based on
the model \pref{sfin} allowed to establish an upper limit $H_{l}$ for
the validity of the linear regime \pref{hlin}
\be
\lb{hlim}
\, H\lesssim 
H_l=0.1 \frac{\Phi_0}{\xi^2}\sqrt{\frac{T-T_{BKT}}{T}}.
\ee
Notice that the above relation can be approximately expressed as the
condition $\xi\gg \ell_B$ for the low-field limit to be applied, where $\ell_B^2=\Phi_0/H$ is the magnetic
length scale.\cite{lesueur_nerst} As $T$ approaches $T_{BKT}$ $\xi$ increases rapidly
and the field $H_{l}$ becomes rapidly smaller than the lowest field
accessible in the standard experimental set-up. This effect can
explain for example the non-linear magnetization effects reported
recently in several measurements in cuprate
superconductors\cite{li_magn_epl05,ong_natphys07,li_nerst_prb10}.
Indeed, the persistence of a non-linear magnetization up to $H\sim
0.01$ T in a wide range of temperatures above $T_{BKT}$ can be a
signature of the rapid decrease of $H_l$ as $T\ra T_{BKT}$, which
does not contradict but eventually support the BKT nature of the SC
fluctuations in these systems. Moreover, since $\xi$ increases as
$\mu$ increases, the extremely low values of $H_l$ measured in Ref.\
\onlinecite{li_magn_epl05} suggest a value of $\mu$ larger than $\mu_{XY}$, in
agreement with the result discussed in Sec. \ref{supdens} based on the
analysis of the superfluid density. On the other hand also the
existence of inhomogeneities can alter the straightforward
manifestation of a linear magnetization above $T_{BKT}$, an issue that
has not been explored yet neither in the context of cuprates nor in
the case of conventional superconductors. Finally, we would like to
mention that even though some theoretical work exists\cite{sondhi_kt}
on the RG approach to the BKT transition at finite magnetic field
based on the Coulomb-Gas analogy, a full analysis of the more general
model \pref{sfin} is still lacking. Such an approach could eventually
improve the estimate \pref{hlim} of the linear regime, based
on a variational calculation that is not expected to capture the correct critical
behavior as the transition is approached.

\section{Conclusions}
\label{conclusions}

It is clear that BKT theory has profoundly changed our understanding of
quasi-2D superconductors and given us a tool to tackle such
challenging and interesting problems. However, more than 40 years after
the original discovery, the occurrence of the BKT transition
in several quasi-2D superconducting materials remains
partly controversial. One can in general identify two possible sources
of discrepancies between theoretical predictions and the current
experimental scenario. From one side, the original formulation was
based on the paradigmatic case of the $XY$ model, that is only one
possible model where the BKT transition occurs. Even though it
correctly reproduces the critical behavior of all the systems belonging
to the same universality class, quantitative discrepancies away from
criticality can be observed in different models. This is the case of
the strong superfluid-stiffness renormalization {\em below} the
transition temperature $T_{BKT}$ in the case of superconducting films
of conventional superconductors, where the vortex-core energy attains
values significantly different from the $XY$ model prediction. From
the other side, emerging new materials and improved experimental
techniques offer new scenarios for the occurrence of the BKT
transition, which coexists with several other phenomena. An example is
provided by the case of cuprate superconductors, that are layered
systems formed by strongly-correlated 2D SC layers. In this case, the
deviations of the vortex-core energy from the $XY$-model value can
eventually lead to a qualitative different behavior of the
superfluid-density jump at the transition or to strong non-linear
field-induced magnetization effects above $T_{BKT}$. In the present
article we reviewed a possible approach to all these issues based on
the sine-Gordon model. Even though this is certainly not a new
approach for the pure 2D case, in the presence of additional relevant
perturbations it provides a very convenient framework to investigate
the BKT physics. Indeed, it allows not only to incorporate easily the
effects of a vortex-core energy value different from the $XY$ model,
but also to describe the coupling to the electromagnetic field in a
clear way, giving a straightforward and elegant description of the
charged superfluid. Finally, we would like to emphasize once more that a
quite interesting issue, that applies equally well to conventional and
unconventional superconductors, is posed by the role of the intrinsic
sample inhomogeneity. Even though we outlined here a kind of
mesoscopic approach to the emergence of spatially inhomogenous SC
properties, a more microscopic approach to the effect of disorder on
the BKT transition would be required, as suggested by
some recent numerical works\cite{meir_epl10,meir_transport_prb11}. The theoretical and
experimental investigation of this issue will certainly offer an other perspective
on the BKT transition in low-dimensional superconductors.

\section*{Acknowledgements}

We thank M. Cazalilla, S. Caprara, A. Caviglia, A. Ho, M. Gabay, S. Gariglio,
M. Grilli, J. Lesueur,  N. Reyren,  P. Raychaudhuri, and J. M. Triscone for
enjoyable collaborations and discussions. This work was supported in
part by the Swiss NSF under MaNEP and Division II.

\bibliographystyle{apsrev}




\end{document}